\documentclass{aa}  
 \pdfoutput=1
 \usepackage{natbib}
\usepackage{graphicx}
\usepackage{txfonts}
\usepackage[breaklinks, colorlinks, urlcolor=Cerulean, citecolor=Cerulean]{hyperref}
\usepackage[usenames,dvipsnames,svgnames,table]{xcolor}

\definecolor{purple}{rgb}{0.62,0.12,0.94}

\newfont{\gwpfont}{cmssq8 scaled 1000}
\newcommand{\rexcess}{{\gwpfont REXCESS}}

\newcommand{\xmm}{XMM-\emph{Newton}}
\newcommand{\Planck}{\emph{Planck}}

\newcommand{\act}{{ACT}}

\newcommand{\chandra}{\emph{Chandra}}

\newcommand{\nika}{\emph{NIKA}}
\newcommand{\mustang}{\emph{MUSTANG}}

\def\be{\begin{equation}}
\def\ee{\end{equation}}
\def\bea{\begin{eqnarray}}
\def\eea{\end{eqnarray}}

\def\xmm{XMM-{\it Newton} }

\def\lesssim{\mathrel{\hbox{\rlap{\hbox{\lower4pt\hbox{$\sim$}}}\hbox{$<$}}}}
\def\gtrsim{\mathrel{\hbox{\rlap{\hbox{\lower4pt\hbox{$\sim$}}}\hbox{$>$}}}}

\providecommand{\sorthelp}[1]{}

\newcommand{\pact}{{PACT}}
\newcommand{\asa}{{\sc PACT31}}
\newcommand{\psa}{{\sc PLCK62}}

\begin{document} 

\defcitealias{arn10}{A10}
\defcitealias{planck2012-V}{P13}
\defcitealias{say13}{S13}
\defcitealias{say16}{S16}

    \title{PACT. II. Pressure profiles of galaxy clusters using \Planck\ and ACT}

   \subtitle{}

     \author{       
         E. Pointecouteau\inst{1},  I. Santiago-Bautista \inst{1,2}, M. Douspis\inst{3}, 
         N. Aghanim\inst{3}, D. Crichton\inst{4}, J.-M. Diego\inst{5},
         G. Hurier\inst{6}, J. Macias-Perez\inst{7},T. A. Marriage\inst{8}, M	. Remazeilles\inst{9}, 
         	C. A. Caretta\inst{2}, H. Bravo-Alfaro \inst{2}}

   \institute{
$^{1}$IRAP, Universit\'e de Toulouse, CNRS, CNES, UPS, (Toulouse), France (\email{etienne.pointecouteau@irap.omp.eu)} \\
$^{2}$Departamento de Astronomía, DCNE, Universidad de Guanajuato, CP 36023, Guanajuato, Gto., Mexico \\
$^{3}$Université Paris-Saclay, CNRS, Institut d'astrophysique spatiale, B\^atiment 121, 91405, Orsay, France.\\
$^{4}$Institute for Particle Physics and Astrophysics, Eidgen\"ossische Technische Hochschule Z\"urich, Wolfgang-Pauli-Str. 27, 8093 Z\"urich, Switzerland\\
$^{5}$Instituto de Física de Cantabria (CSIC-UC), Avda. Los Castros s/n, 39005 Santander, Spain\\
$^{6}$Centro de Estudios de Física del Cosmos de Aragón (CEFCA), Plaza de San Juan, 1, Planta 2, 44001 Teruel, Spain\\
$^{7}$LPSC, Université Joseph Fourier Grenoble 1, CNRS/IN2P3, Institut National Polytechnique de Grenoble,53 Av. des Martyrs, 38026 Grenoble, France\\
$^{8}$Dept. of Physics and Astronomy, Johns Hopkins University, 3400 N. Charles St., Baltimore, MD 21218, USA\\
$^{9}$Jodrell Bank Centre for Astrophysics, School of Physics and Astronomy, The University of Manchester,Manchester M13 9PL, UK\\
             }

\date{Submitted December 23, 2020; accepted May 10, 2021}

  \abstract
  {The pressure of hot gas in groups and clusters of galaxies is a key physical quantity, which is directly linked to the total mass of the halo and several other thermodynamical properties. In the wake of previous observational works on the hot gas pressure distribution in massive halos, we have investigated a sample of 31 clusters detected in both the \Planck\ and Atacama Cosmology Telescope (ACT), MBAC surveys. We made use of an optimised Sunyaev-Zeldovich (SZ) map reconstructed from the two data sets and tailored for the detection of the SZ effect, taking advantage of both \Planck\ coverage of large scales  and the \act\ higher spatial resolution. Our average pressure profile covers a radial range going from $0.04 \times R_\textrm{500}$ in the central parts to $2.5 \times R_\textrm{500}$ in the outskirts. In this way, it improves upon previous pressure-profile reconstruction based on SZ measurements. It is compatible, as well as competitive, with constraints derived from joint X-ray and  SZ analysis. This work demonstrates  the possibilities offered by large sky surveys of the SZ effect with multiple experiments with different spatial resolutions and spectral coverages, such as \act\ and \Planck.
}
   \keywords{Galaxy clusters --
   	intracluster medium --
   	submilimiter
               }
\titlerunning{The pressure profile of PACT clusters}
\authorrunning{AA et al.}
   \maketitle
%

\section{Introduction}

The thermal pressure from the hot gas in massive dark matter halos is the main force preventing the gravitational collapse of the gas towards their centre. Thermal pressure is fuelled by the infall of baryonic matter into the potential wells of groups and clusters of galaxies,  inducing gravitational heating of the gas to temperatures of $\sim 10^{7}-10^{8}$~K, that is $\sim1-10$~keV. The gas pressure is a key quantity for the physical characterisation of these systems and can be investigated either from the plasma X-ray emission \citep{sar88} or from its interaction with the cosmic microwave background (CMB), that is from the Sunyaev-Zeldovich effect \citep[SZ effect hereafter][]{sun72}. The former provides an indirect reconstruction of the pressure via the measurement of the gas density, $n_e$, and temperature, $T$, from the X-ray surface brightness, $S_X\propto n_e^2 \sqrt{T}\exp(-E/T)$, where $E$ is the energy. The gas pressure is derived as $P = n_e \times T$. Conversely, the SZ effect is a direct probe of the gas pressure, since the integrated SZ flux (quantified hereafter by the integrated Comptonisation parameter, $Y$) over the clusters directly links to the pressure as $Y \propto\int_V PdV$.

As expected from the simplest spherical collapse scenario, the population of groups and clusters of galaxies manifest several properties of similarity \citep{kai95, ber98}. This behaviour is observed via their global thermodynamical properties \citep[e.g.][]{gio13}  as well as for their internal distribution \citep[e.g.][]{pra19}. The gas thermal pressure is a remarkable example of this self-similar behaviour. The integrated pressure over the volume of the cluster, that is the SZ flux, has been shown to be an excellent proxy of the total gas content, thus of the total mass of the halo. Indeed the thermal pressure is mildly affected by non-gravitational physics (AGN feedback, radiation cooling, etc) relative to other proxies \citep[e.g. X-ray total luminosity, see][for recent reviews]{pra19,mro19}. 

The advances made by SZ observations in the last two decades have allowed a precise measurement of the integrated pressure over statistically significant samples \citep{planck2011-5.2a, planck2012-III, cza14, ben16,die19},  demonstrating the coherent view of the gas content of galaxy clusters between X-ray and millimetre measurements. Increases in the spectral coverage, spatial resolution, and sensitivity of SZ observations have also improved constraints on the pressure distribution over the whole volume of clusters \citep{pla10, planck2012-V, say13, eck13}.  Both theory and numerical simulations of structure formation have provided a successful description of the gas behaviour under the influence of the key physical processes governing the intra-cluster medium \citep[ICM,][]{nag07, bat10}.  One outcome of these simulations is the generalisation of the  \citet{nfw96} profile (gNFW) for the distribution of dark matter derived from early numerical simulations, to the one for the gas pressure by \citet{nag07}:

\begin{equation}
\mathbb{P}(x)=\frac{P_0}{(c_\textrm{500}\, x)^\gamma [1+(c_\textrm{500}\, x)^\alpha]^{(\beta-\gamma)/\alpha}}
\label{e:gnfw}
\end{equation}

\noindent where $x=r/r_s$ and $r_s=r_\textrm{500}/c_\textrm{500}$. The quantities $r_\textrm{500}$ and $c_\textrm{500}$ are the characteristic radius and the concentration corresponding to a density contrast of 500 times the critical density of the Universe at the cluster redshift. The exponents $\gamma$, $\beta$, and $\alpha$ are the inner, external, and transition slopes (at $r_s$) of the profile, respectively.  The gNFW profile thus provides a simple parametric description, which can be tested against observational constraints \citep[e.g.][]{arn10,planck2012-V, eck13, ada15, ada16, say16, rom17, bou17, rup18}. The afore-cited works have found a very good agreement between the gNFW model and the observed pressure distribution in X-ray or SZ, at least within the central part of the galaxy clusters. Beyond $R_\textrm{500}$, the observational constraints  mainly come from SZ observations and show an average agreement with the gNFW profile, although with some scatter \citep[e.g.][]{say16, ghi18}. These variations are potentially linked to disparate samples  providing a non-homogeneous sampling of the cluster population. They might also be the imprint of intrinsic variations in the outskirts across the population of massive halos due to the complex physics governing the virialisation of the gas (e.g. shocks, turbulent and bulk motions of the gas, and complex accretion from the larger surroundings leading, for instance, to inhomogeneities and substructures \citealt{sim19,wal19}).

In this paper, we pursue this observational investigation of the pressure distribution in clusters on the basis of the  combination of the \Planck\ and \act\ maps by \citet{agh19} \citep[see also the work by][]{mad20}. The resulting reconstructed SZ map  is optimised for the detection of the SZ signal with its  $\sim 1.5$~arcmin spatial resolution and a tightly controlled noise. From the footprints of the two \act-MBAC survey strips \citep{dun13}, we have assembled a sample of massive clusters of galaxies detected by both \Planck\ and \act in order to investigate their thermal pressure distribution on the basis of the SZ observations only. This study is thus a test and pathfinder case for future work on larger survey areas, for example, the 18\,000 sq. deg. jointly covered by \act\ and \Planck\ \citep{aio20}. 

In the following section we briefly describe the \pact\ SZ map and the cluster sample defined for this work. In Sec.~\ref{s:rec}, we recall some  basics on the SZ effect and the methodology adopted to reconstruct the pressure distribution in our clusters. Sec.~\ref{s:val} presents the  validation procedure based on the comparison to the \Planck\ data, and the related results of the various validation steps. The $y$- and pressure profiles are given in Sec.~\ref{s:pr}, before discussing the outcome of our work in Sec.~\ref{s:dis}

Throughout this paper, we assume a $\Lambda$CDM cosmology with H$_0~=~70~$km~s$^{-1}$~Mpc$^{-1}$, $\Omega_M~=~0.3$, and $\Omega_\Lambda~=~0.7$.

\begin{table*}[!t]
\centering          
\caption{Properties of the \asa\ sample}
{\small \begin{tabular}[]{c c l c c c c c c c c}
		\hline\hline                 
Name & PSZ2 Name  & \act\ Name& $\alpha$ & $\delta$ & $z$ & $M_\textrm{500}$ & $R_\textrm{500}$&    \multicolumn{3}{c}{S/N} \\
     &    &  & [Deg] & [Deg] &      & $10^{14}$~M$_\odot$  & [kpc] & \act\ & PSZ1 & PSZ2  \\
\hline
C00 & PSZ2 G101.55-59.03 & ACT-CL J0008.1+0201 &   2.042 &    2.020 & 0.3651 & 5.72 & 1113 & 6.11 & 4.80 & 4.70 \\
C01 & PSZ2 G119.30-64.68 & ACT-CL J0045.2-0152 &  11.305 &   -1.883 & 0.5450 & 6.37 & 1073 & 5.08 & -- & 7.50 \\
C03 & PSZ2 G130.21-62.60 & ACT-CL J0104.8+0002 &  16.219 &    0.049 & 0.2770 & 5.71 & 1147 & 6.19 & 4.74 & 4.30 \\
C05 & PSZ2 G153.00-58.26 & ACT-CL J0152.7+0100 &  28.176 &    1.006& 0.2270 & 5.04 & 1120 & 5.64 & 4.54 & 9.00 \\
C08 & PSZ2 G172.98-53.55 & ACT-CL J0239.8-0134 &  39.972 &   -1.576 & 0.3730 & 7.64 & 1219 & 7.62 & 6.15 & 8.80 \\
C10 & PSZ2 G173.90-51.89 & ACT-CL J0245.8-0042 &  41.465 &   -0.701 & 0.1790 & 3.74 & 1032 & 4.56 & -- & 4.10 \\
C12 & PSZ2 G181.44-44.76 & ACT-CL J0320.4+0032 &  50.124 &    0.540& 0.3939 & 5.14 & 1064 & 4.64 & 5.09 & 4.90 \\
C13 & PSZ1 G184.23-44.26 & ACT-CL J0326.8-0043 & 51.708 &   -0.731 & 0.4500 & 6.65 & 1131 & -- & 4.78 & 9.10 \\
C23 & PSZ2 G044.58-20.46 & ACT-CL J2025.2+0030 & 306.301 &    0.513 & 0.2746 & 5.65 & 1117 & 6.59 & 5.02 & 6.40 \\
C24 & PSZ2 G048.91-25.55 & ACT-CL J2050.7+0123 &312.681 &    1.386 & 0.3334 & 5.44 & 1106 & 5.56 & 5.01 & 7.40 \\
C25 & PSZ2 G049.80-25.16 & ACT-CL J2051.1+0215 & 312.789 &    2.263 & 0.3211 & 6.39 & 1172 & 7.59 & 6.03 & 5.20 \\
C26 & PSZ2 G050.06-27.32 & ACT-CL J2058.8+0123 & 314.723 &    1.384& 0.3340 & 6.60 & 1185 & 6.78 & 5.51 & 8.30 \\
C27 & PSZ2 G054.95-33.39 & ACT-CL J2128.4+0135 & 322.104 &    1.600 & 0.3920 & 7.32 & 1197 & 7.30 & 5.89 & 7.30 \\
C29 & PSZ2 G053.44-36.25 & ACT-CL J2135.1-0102 & 323.791 &   -1.040& 0.3300 & 7.45 & 1229 & 8.50 & 7.78 & 4.10 \\
C30 & PSZ2 G055.95-34.89 & ACT-CL J2135.2+0125 & 323.815 &    1.425& 0.2310 & 6.73 & 1233 & 9.54 & 9.09 & 9.30 \\
C31 & PSZ2 G059.81-39.09 & ACT-CL J2156.1+0123 & 329.041 &    1.386& 0.2240 & 4.99 & 1118 & 6.77 & 5.13 & 6.00 \\
C32 & PSZ1 G080.66-57.87 & ACT-CL J2327.4-0204 & 351.866 &   -2.078 & 0.7050 & 8.30 & 1100 & -- & 6.37 & 13.1 \\
C33 & PSZ2 G087.03-57.37 & ACT-CL J2337.6+0016 & 354.416 &    0.269& 0.2779 & 7.33 & 1248 & 11.9 & 7.50 & 8.20 \\
C06 & PSZ2 G276.75-59.82 & ACT-CL J0217-5245&  34.296 &  -52.756 & 0.3432 & 4.48 & 1033 & 5.44 & -- & 4.10 \\
C07 & PSZ2 G270.93-58.78 & ACT-CL J0235-5121&  38.967 &  -51.354 & 0.2780 & 5.95 & 1163 & 8.96 & 6.03 & 6.20 \\
C09 & PSZ2 G271.53-56.57 & ACT-CL J0245-5302&  41.388 &  -53.034& 0.3000 & 6.77 & 1204 & 10.4 & 7.75 & 9.10 \\
C11 & PSZ2 G263.03-56.19 & ACT-CL J0304-4921 & 46.062 &  -49.362 & 0.3920 & 4.70 & 1030 & 4.66 & 4.64 & 3.90 \\
C14 & PSZ2 G264.60-51.07 & ACT-CL J0330-5227 &  52.725 &  -52.468 & 0.4400 & 6.93 & 1149 & 10.8 & 7.83 & 6.10 \\
C15 & PSZ2 G262.73-40.92 & ACT-CL J0438-5419 & 69.579 &  -54.318 & 0.4210 & 7.46 & 1188 & 12.7 & 10.9 & 8.00 \\
C16 & PSZ2 G261.28-36.47 & ACT-CL J0509-5341 & 77.338 &  -53.701 & 0.4607 & 4.18 & 964 & 5.07 & -- & 4.80 \\
C17 & PSZ2 G262.27-35.38 & ACT-CL J0516-5430 & 79.125 &  -54.508& 0.2952 & 8.76 & 1315 & 22.9 & 18.7 & 4.60 \\
C18 & PSZ2 G260.63-28.94 & ACT-CL J0559-5249 &  89.929 &  -52.820 & 0.6009 & 5.96 & 1023 & 7.75 & 4.99 & 5.10 \\
C19 & PSZ2 G263.14-23.41 & ACT-CL J0638-5358 & 99.692 &  -53.979 & 0.2266 & 6.83 & 1242 & 12.8 & 10.8 & 10.0 \\
C20 & PSZ2 G263.68-22.55 & ACT-CL J0645-5413 &101.375 &  -54.228& 0.1644 & 7.96 & 1333 & 21.7 & 17.4 & 7.10 \\
C21 & PSZ2 G266.04-21.25 & ACT-CL J0658-5557 & 104.625 &  -55.951& 0.2965 & 13.1 & 1503 & 28.4 & 20.5 & 11.5 \\
C22 & PSZ2 G265.86-19.93 & ACT-CL J0707-5522 & 106.804 &  -55.380 & 0.2960 & 4.64 & 1063 & 5.77 & 4.88 & 3.30 \\

\hline
\end{tabular}
}
\label{t:sa}
\end{table*}

\section{The \Planck\ and \act\ sample and SZ data}
\label{s:pacs}

\subsection{The joint SZ map}
\label{s:map}
To extract the SZ signal for each individual cluster of our sample, we employed the joint \Planck\ and \act\ SZ map \citep[hereafter called the \pact\ map,][]{agh19}, that is a $y$-map reconstructed from the linear combination of the \Planck\ \citep{planck2014-a09} and \act\  frequency maps \citep{dun13}. While \Planck\ is an all-sky survey, the \act\ map is constituted of two strip maps, an equatorial and a southern one. 
This reconstruction is performed making use of an internal linear combination (ILC) method, MILCA \citep[Modified Internal Linear Combination Algorithm][]{hur13}. 
{Such methods perform an optimal combination of frequency maps (from a single instrument or several instruments) for the reconstruction of a targeted frequency-dependent signal, that is the SZ effect in our case. They account for the intrinsic resolution and noise  (instrumental and astrophysical) of each frequency map included in the combination}.

As originally shown in \citet{rem13}, this combination takes advantage of the \Planck\ frequency coverage and \act\ spatial resolution. Also, \Planck\ has the unique ability to provide the large spatial scales, which are excluded from the \act\ signal after spatial filtering needed to reduce the impact of atmospheric brightness fluctuations. Hence the final \pact\ $y$-map inherits the spatial resolution of the \act\ survey, which  can be well approximated by a $1.4$~arcmin FWHM Gaussian, out to the outskirts of low-$z$ systems provided by \Planck.  We refer to the first \pact\ paper, Sec.~3.2 for a detailed description of the $y$-map reconstruction and  its characterisation \citep{agh19}. The MILCA $y$-map  and its associated noise map are provided in units of the dimensionless Comptonisation parameter. Integrated measured SZ fluxes are expressed in units of  arcmin$^2$ and the SZ luminosities in Mpc$^2$. 
We also made use, for validation purposes (see Sec.~\ref{s:val}), of the \Planck\ all-sky $y$-map. We used the public MILCA $y$-map \citep{planck2014-a28} which has an angular resolution of 10~arcmin FWHM. 
We also used a second (non-public) version of the MILCA $y$-map reconstructed at 7~arcmin FWHM. This map has been used for the extraction of the \Planck\ SZ signal by the X-COP collaboration \citep{tch16,eck17,ghi18}. 

\subsection{The cluster samples}
\label{s:sa}
We have defined our samples from SZ catalogues of galaxy clusters obtained from the \Planck\ \citep[][ESZ, PSZ1, PSZ2, respectively]{planck2011-5.1a,planck2013-p05a,planck2014-a36} and \act\ \citep{has13,hil18} surveys. A total of 119 clusters are detected within the \act\ footprint by either instrument. Here we focus on the  34 joint detections.  We exclude three sources  partially covered at the edges of \act\ footprint or falling into the mask of  \Planck\ point sources  (used to construct the \pact\ $y$-map). Our final sample is thus comprised of 31 clusters of galaxies, hereafter referred as \asa, with 18 sources distributed in the equatorial strip and 13 in the southern one.
By construction our sample is neither representative nor complete. The clusters range between 0.16 and 0.70 in redshift and  from $3.7\times10^{14}$ to  $1.3\times10^{15}$~M$_\odot$ in $M_\textrm{500}$, where $M_{500}$ is the total mass contained within $R_\textrm{500}$. The mean signal-to-noise ratio (S/N) of the sample is 6.0 in the \Planck\ catalogues and 6.3 in \act's. The range of angular size over the sky is $2.5<\theta_\textrm{500}<7.9$~arcmin, with a mean value of $4.2\pm1.1$~arcmin.  

We also consider the sample of 62 massive local clusters used in \citet[][P13 hereafter]{planck2012-V} to derive the pressure profile of the hot intra-cluster gas from the first \Planck\ all-sky survey. Here the \Planck\ maps used to derive the $y$-map (see above) are those from the  all-sky survey \citep[i.e. 2015 data release][]{planck2014-a09}. We therefore use this sample, hereafter \psa, in order to compare to the profiles derived from our  \pact\ map (see Sec.~\ref{s:val}) . From the  second \Planck\ catalogue, PSZ2, the redshift of the \psa\ sample ranges between 0.04 to 0.44 and the  S/N  from 7 to 49.
The covered  mass interval is   $2.4\times 10^{14} < M_\textrm{500}< 2.0\times 10^{15}$~M$_\odot$, for an angular size one of $3.7<\theta_\textrm{500}<22.8$~arcmin with a mean value of $9.8\pm5.4$~arcmin.

\begin{figure*}[!t]
   \centering
   \includegraphics[width=0.33\textwidth]{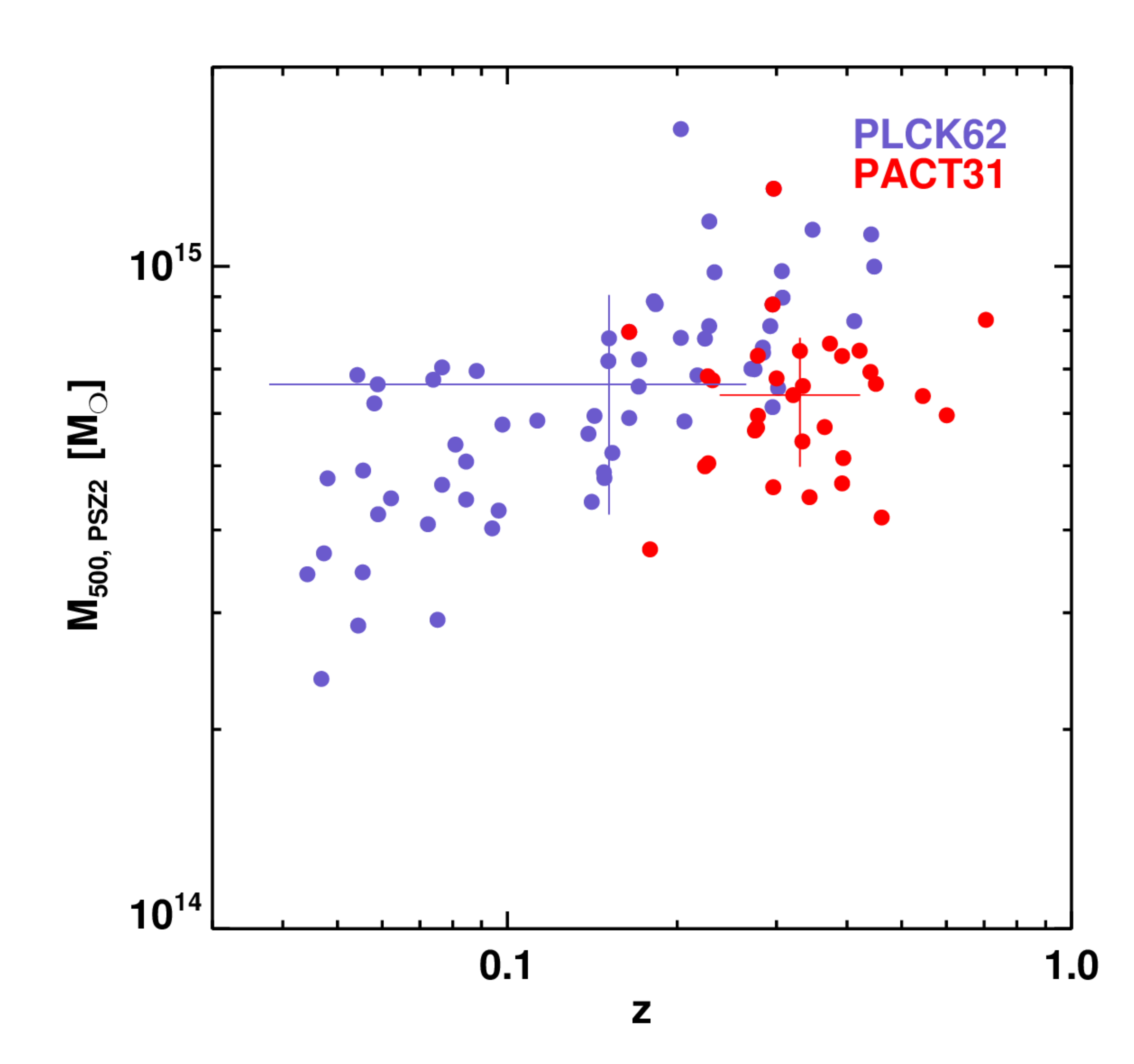}%
   \includegraphics[width=0.33\textwidth]{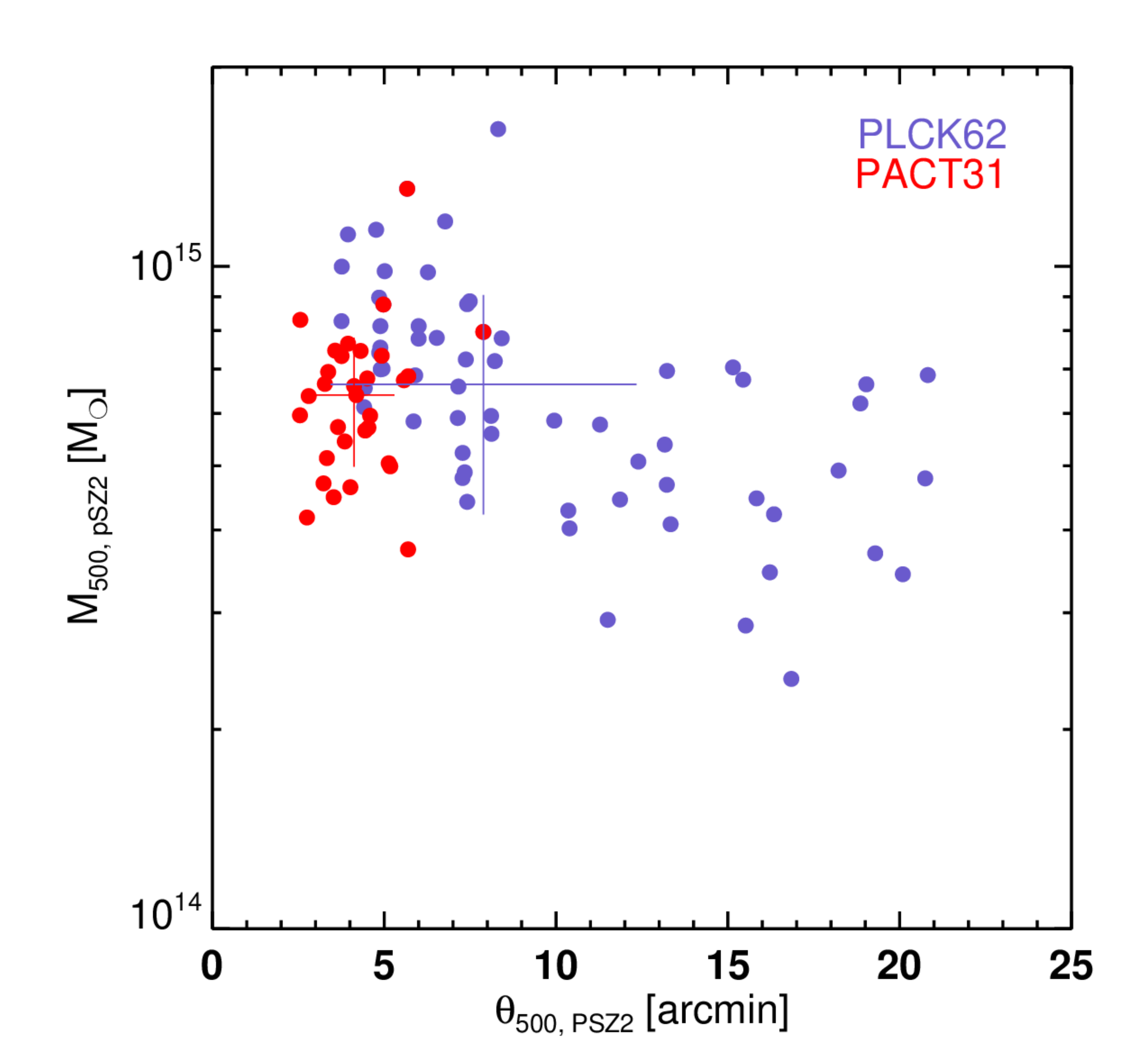}
   \includegraphics[width=0.33\textwidth]{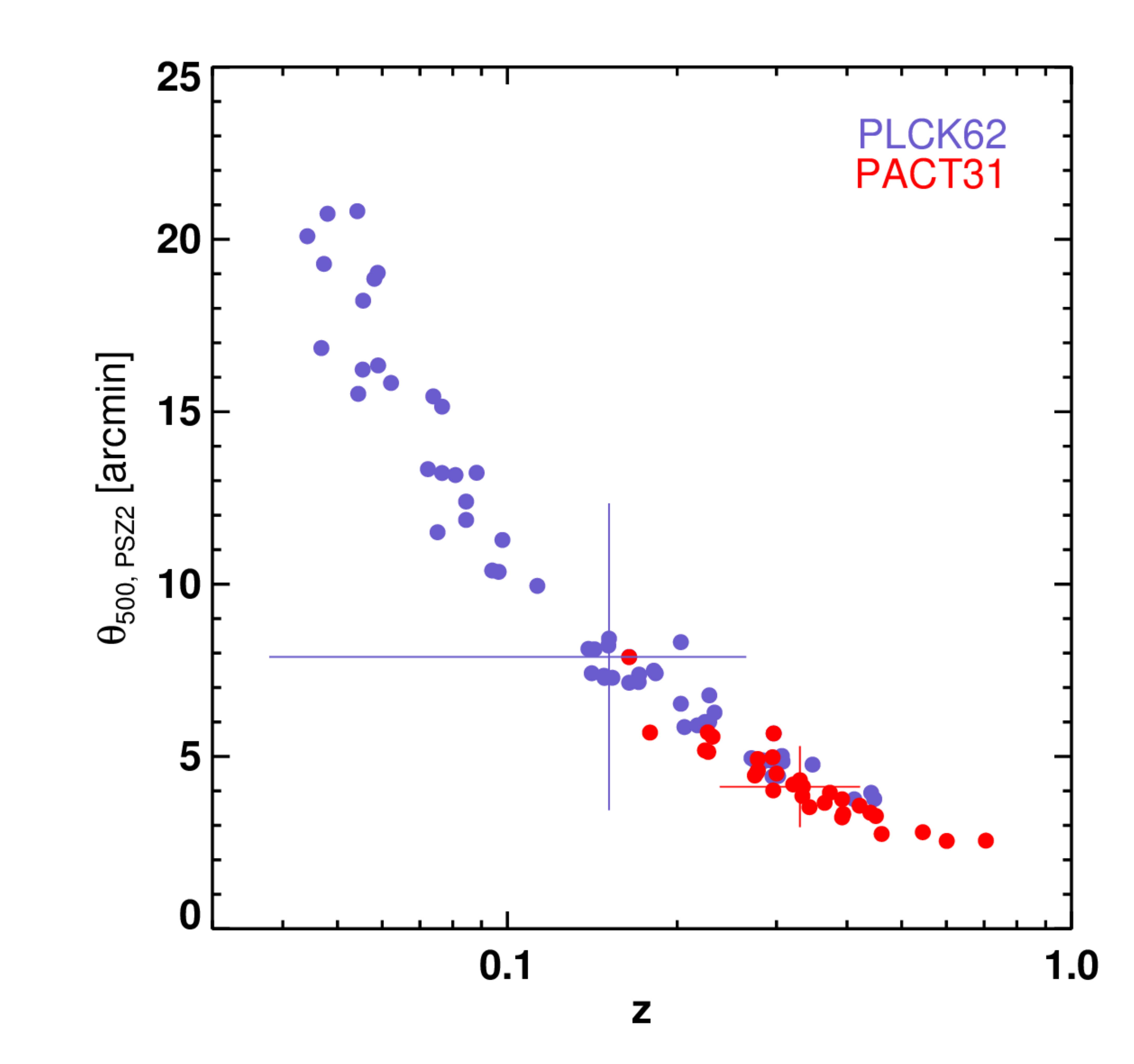}
      \caption{Distribution of the {\sc PACT31} (red dots) and {\sc PLCK62} (blue dots) in the  $M_\textrm{500}-z$ \emph{(left)}, $M_\textrm{500}-\theta_\textrm{500}$ \emph{(middle)} and $\theta_\textrm{500}-z$ \emph{(right)}. Crosses picture the median value and associated maximum absolute deviation over each sample. Values for the redshift, masses and angular radii are taken from the PSZ2 catalogue \citep{planck2014-a36}.}
         \label{f:sa}
   \end{figure*}

Masses for the \psa\ sample are $M_\textrm{500}$ hydrostatic masses from \citetalias{planck2012-V}, derived from the \xmm observations. For the \asa\ sample, we make use of the $M_\textrm{500}$ values provided in the PSZ2 catalogue. 
The latest are derived from an $Y_\textrm{500}-M_\textrm{500}$ relation calibrated with \Planck\ SZ integrated fluxes and \xmm hydrostatic masses \citep{planck2013-p15,planck2014-a30}. We checked the consistency between the \citetalias{planck2012-V} hydrostatic mass and the PSZ2 masses for the \psa\ sample, finding a mean ratio of $0.98\pm0.11$.  
We provide, in Fig.~\ref{f:sa}, the distribution of the two samples in the $M_\textrm{500}-z$, $M_\textrm{500}-\theta_\textrm{500}$ and  $\theta_\textrm{500}-z$ plans.

\section{Computation of the $y$ and pressure profile}
\label{s:rec}

To recover the pressure profiles from the SZ $y$-map, we strictly followed the method described and used by  \citetalias{planck2012-V}.
The SZ flux is the product of the SZ spectrum integrated over a frequency band and the integrated Compton parameter, $Y$, within a given solid angle, $\Omega$. The latter is proportional to the thermal pressure of the ICM gas integrated over the line of sight:

\begin{equation}
Y(\Omega) =   \frac{\sigma_{\rm T}}{m_{\rm e} c^2} \int_\Omega d\Omega  \int_{los} P(l)\, dl .
\label{e:sz}
\end{equation}

The effect of the weakly relativistic velocities of the electrons as a function of the gas temperature on the SZ spectrum shape \cite[e.g.][]{poi98} are neglected in the  $y$-map reconstruction as accounting for these effects would require a priori knowledge on the gas temperature distribution across the cluster. 

%
From the $y$-map and error map, we extracted patches from the $20\times \theta_\textrm{500}$ side centred on the \act\ cluster position. We chose a pixel size defined in constant units of $\theta_\textrm{500}$, which is thus common to all clusters in our sample (i.e. all 31 patches have different angular pixel sizes). The pixel size also implies a sampling of the \pact\ PSF and therefore a possible oversampling of the $y$-map pixel (Sec.~\ref{s:val}). We take into account and propagate the correlations induced by this oversampling factor.

We compute an azimuthal $y$-profile from the map. The value in each radial bin is obtained from the mean of the pixel values  it encompasses. 
A background offset is estimated from radii greater than $5\times R_\textrm{500}$ and is subtracted. 
We masked point sources in a two step process: (i) obvious positive or negative\footnote{Negative sources are due to negative coefficients in the linear combination of  frequency maps.} sources are masked manually,
and (ii) a pixel clipping is applied on the map with a $2.5\sigma$ criterium with respect to the mean map flux outside a radius of $5\times R_\textrm{500}$.
 To account for the correlation introduced by the aforementioned sampling and intrinsic correlated noise of the \pact\ $y$-map, we computed the covariance matrix associated to the $y$-profile.
 To do so, we estimated the power spectrum of the noise (astrophysics, instrument, systematics) in the region surrounding the cluster  ($\theta>5\times R_\textrm{500}$). We drew a thousand realisations of the noise patch, applied the same profile extraction as for the $y$-patches, and derived the covariance matrix {as $C=N_{n,m}^\textrm{T}N_{n,m}$ (where $N_{n,m}$ is a matrix of $n$ points per profiles $\times$ $m$ simulated noise profiles).}

The 2D $y$-profiles are binned  to maximise S/N out to the largest possible radius. To derive the 3D pressure profiles, we assumed spherical symmetry of our clusters and applied a regularised PSF deconvolution and geometrical deprojection algorithm, adapting the method described in \citet{cro06} for X-ray surface brightness profiles. Errors encoded in the 2D $y$-profile covariance matrix were propagated to the 3D pressure profile {on the basis of the matrix's Choleski decomposition (assuming correlated Gaussian noise), from which $10000$ random realisations of the $y$-profile are drawn. Each of these realisations is deconvolved} from the PSF and deprojected individually. 
In the process, the dispersion in flux in each radial bin, derived from $y$-map, is used as a weight for each point of the $y$-profile.
{The covariance matrix of the pressure profile, $C_P$, is derived from the combination of $l=10000$ realisations of the pressure profile,  $C_P=N_{n,l}^TN_{n,l}$ (where $N_{n,l}$ is a matrix of $n$ points per profiles $\times$ $m$ simulated noise profiles)}. Both profile and covariance matrix are then scaled in physical units.

We followed the stacking procedure described in Sec.~4.3.2, Eq.~14 of P13 to stack the $y$-profiles. We recall here that the output stacked profile and associated covariance matrix are:
\begin{equation}
\tilde{y}=\frac{1}{n}\sum_i^n{\frac{y_i}{\Phi_i}}\;\;\;\;\;\; {\rm 
and}\;\;\;\;\;\; \widetilde{C}=\frac{1}{n^2}\sum_i^n{\frac{C_i}{\Phi_i^2}}, \\
\label{e:norm}
\end{equation}
with $\Phi_i=Y_{\textrm{500},i}/R_{\textrm{500},i}^2$  for the $i^{th}$ cluster, $n$ the number of clusters in the sample and $y_i$ and $C_i$ the  $y$-profile and associated covariance matrix for the $i^{th}$ cluster. 
The staked pressure profile and its covariance matrix are derived in a similar way, with $\Phi_i=P_{\textrm{500},i}\times f(M_i)$ for the $i^{th}$ cluster. $P_\textrm{500}$ is the pressure integrated within $R_\textrm{500}$ and $f(M)=(M_\textrm{500}/3\times10^{14} h_{70}^{-1} M_\odot)^{0.12}$.

   \begin{figure*}[!t]
   \centering
   \includegraphics[width=0.33\textwidth]{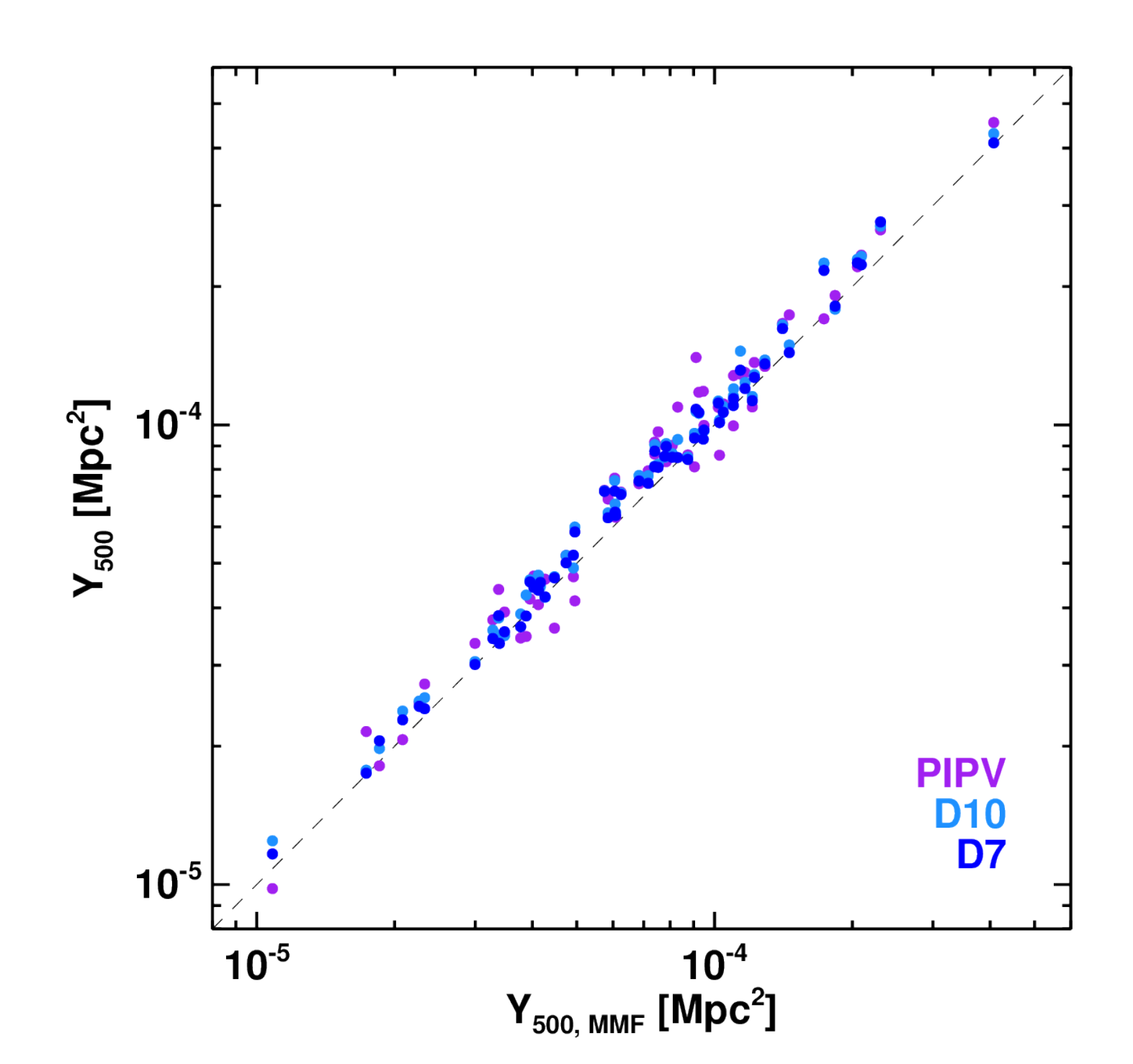}%
   \includegraphics[width=0.33\textwidth]{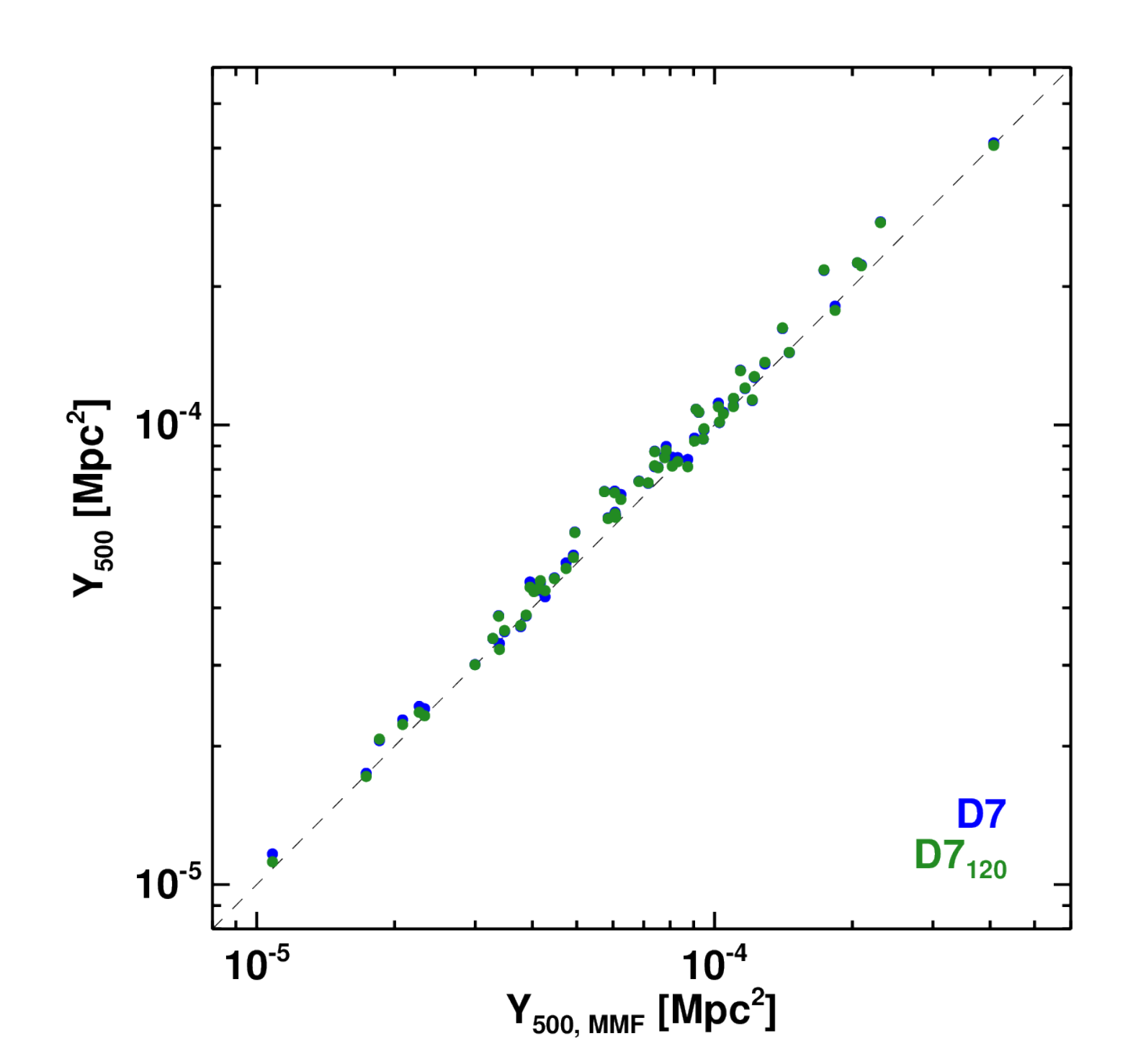}%
   \includegraphics[width=0.33\textwidth]{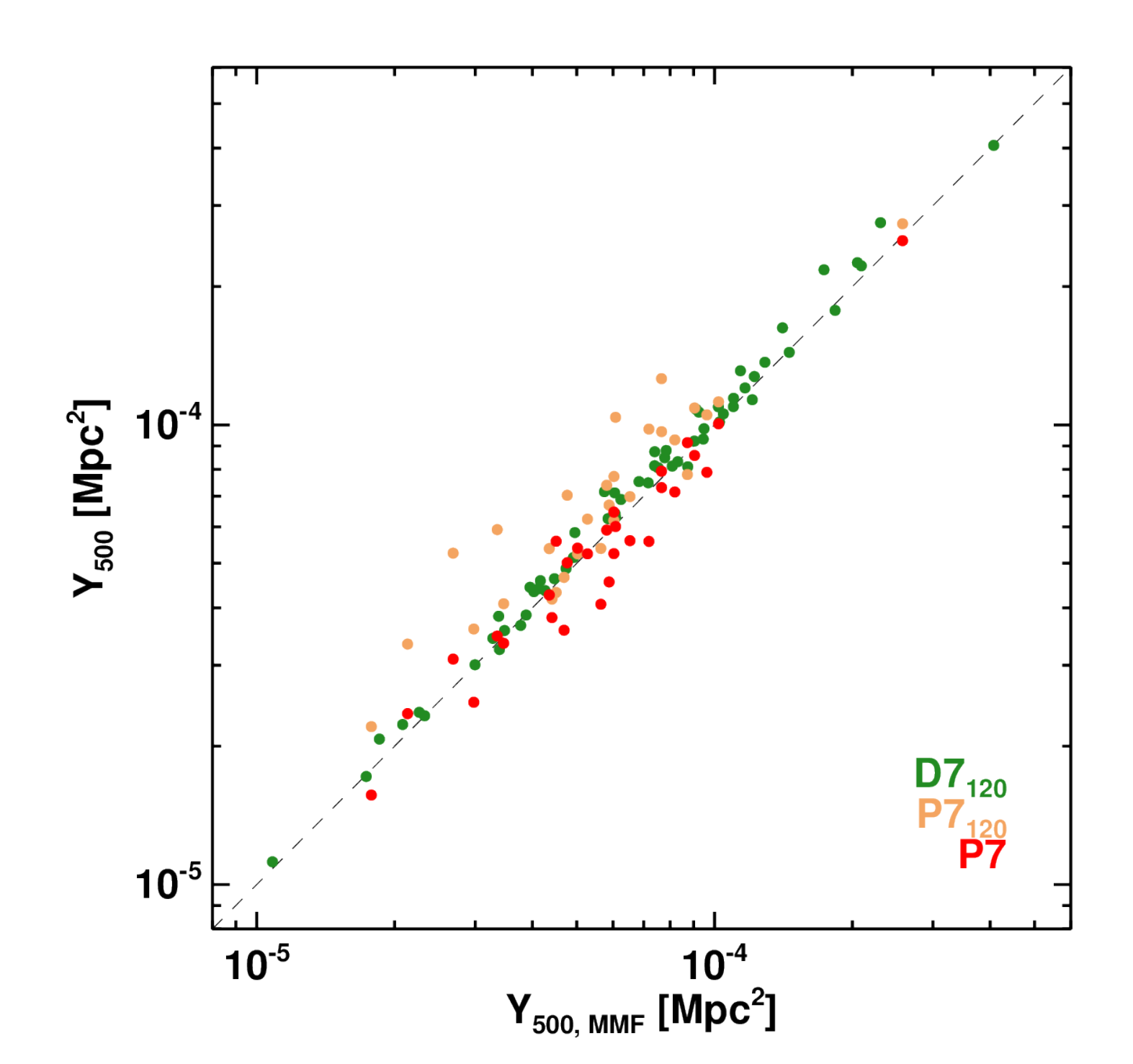}\\%
   \includegraphics[width=0.33\textwidth]{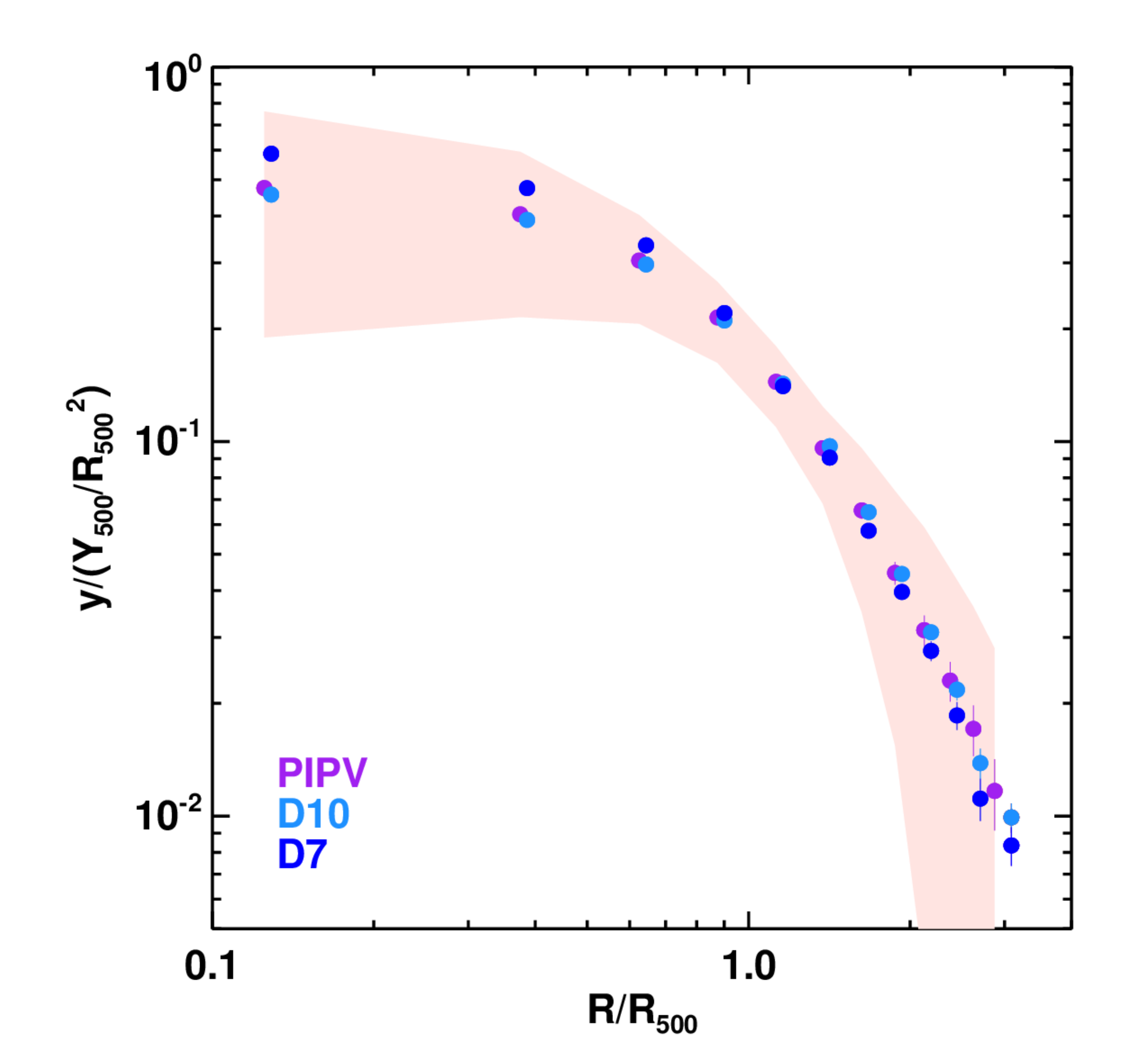}%
   \includegraphics[width=0.33\textwidth]{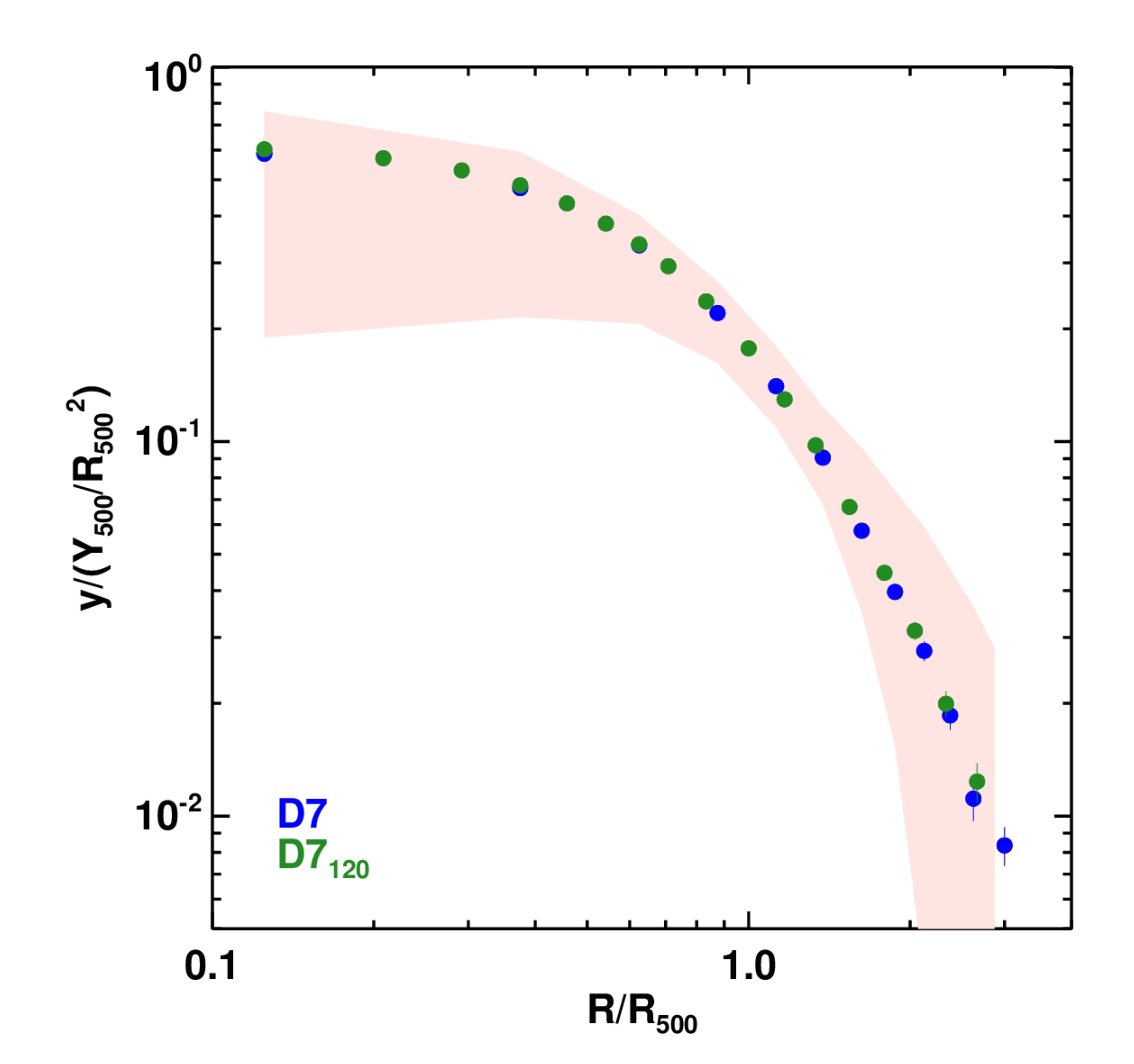}%
   \includegraphics[width=0.33\textwidth]{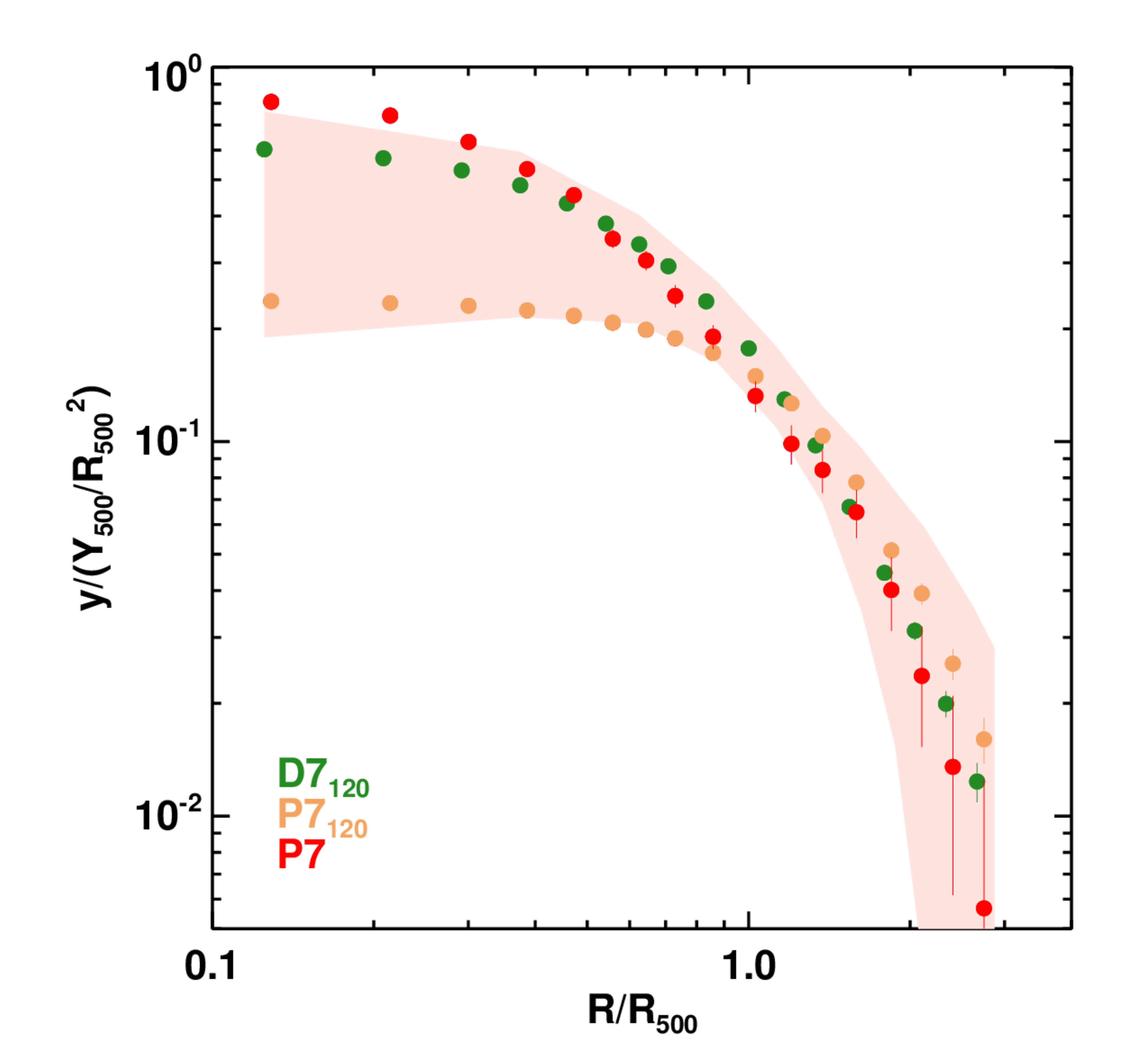}%
    \caption{\emph{(top-row)} One-to-one comparison of the integrated SZ flux, $Y_\textrm{500}$,  using as reference $x$-axis values fluxes computed from the \pact\ maps as presented in \citet{agh19}. \emph{(bottom-row)} Average stacked $y$ profiles in units of $R_\textrm{500}$ and $Y_\textrm{500}/R_\textrm{500}^2$. Columns present the various steps of validation:  \emph{(left-column)} Check of \Planck\ dataset version and $y$ map resolution. \emph{(middle-column)} Check of the impact of the radial sampling factor $\delta_r$. \emph{(right-column)} Comparison of profiles derived from \Planck\ and \pact\ maps and samples. Labels in  the legends are for the various test configurations as defined in Table~\ref{t:val} and discussed Sec.~\ref{s:val}. For the three profile plots,  the red shaded envelop is the same and corresponds to the $1\sigma$ dispersion across this same sample as published by \citetalias{planck2012-V} (left panel of their figure~3). We also adopted their radial range. The error bars correspond to the square root of the diagonal of the covariance matrix and therefore bear a certain degree of correlation between points.}
         \label{f:val}
   \end{figure*}

%
%
\section{Validation of the \pact\ profiles}
\label{s:val}

To fully assess our $y$ and pressure profiles derived from the \pact\ maps, we proceeded  through the following validation procedure which relies on two quantities: the SZ $y$-profiles and the  SZ flux $Y_\textrm{500}$ (integrated within the  radius $R_\textrm{500}$  as projected on the sky). Radial profiles, $y(\theta)$, are derived from the $y$-map together with their correlation matrix as described in the previous section. Each integrated SZ flux is derived from the corresponding $y$-profile, assuming a universal pressure profile \citep[][A10 hereafter]{arn10}, 
{ which is projected and convolved by the \pact\ PSF into a $y$-profile model, $M$. Folding in the covariance matrix, $C$, of the $y$-profile,  the SZ flux, $Y_\textrm{500}$, is obtained from the optimal solution minimising the chi-square~: 
$Y_\textrm{500}=\sigma^2_{Y_\textrm{500}} M^T C^{-1} y(\theta)$, with  $\sigma_{Y_\textrm{500}}^2 = (M^T C^{-1} M)^{-1}$.}

\begin{table}[!b]
\centering          
     \caption[]{Definition of setups for the validation procedure}
         \label{t:val}
\begin{tabular}{lllrc}
\hline
\hline
 Name & Sample & $y$-map & FWHM & $\Delta r/R_\textrm{500}$   \\
\hline
 PIPV	 & \psa\ & PC-internal & 10~arcmin & 0.25  \\
 D10     & \psa\ & DR2015 & 10~arcmin & 0.25  \\
 D7       & \psa\ &  DR2015 & 7~arcmin & 0.25 \\
 D7$_{120}$       & \psa\ & DR2015 &   7~arcmin & 0.08  \\
 P7$_{120}$     & \asa\ & DR2015 &  7~arcmin & 0.08 \\
 P7     & \asa\ & \pact\ &  1.4~arcmin & 0.08  \\
\hline
\end{tabular}
\end{table}

\subsection{Working method and setup}

We computed $y(\theta)$ and $Y_\textrm{500}$ over the public and non-public \Planck\ MILCA $y$-maps, with 10 and 7~arcmin FWHM, respectively, for the \psa\ sample (i.e. {named configurations D10 and D7, respectively}), and over the latter only for the \asa\ sample (i.e. configuration P7$_{120}$). 

We first used the same setup adopted by \citetalias{planck2012-V} for the radial sampling, with $\delta_r=\Delta\theta/\theta_\textrm{500}=\Delta r/R_\textrm{500}=0.25$. We also later on used a value of  $\delta_r=0.08$, a higher resolution sampling more appropriate for the \pact\ PSF. This value is derived  in order to properly sample the \pact\ PSF accounting for the angular extension of our clusters (see Table~\ref{t:sa}). The PSF sampling imposed the need for $\delta_r < \textrm{PSF}/2=0.7$~arcmin. As we extracted our profiles over a regular grid in units of $R_\textrm{500}$ out to a value of 10, the number of radial bins is fixed by the cluster in our sample with the largest angular extension, that is C20 with $\theta_\textrm{500}=7.9$~arcmin. This led to a minimum of 112 bins with $10\times\theta_\textrm{500}$, which we rounded up to 120 points. The PSF sampling thus complies with the Nyquist-Shannon criteria for all our objects, with the PSF oversampling rate increasing for clusters with smaller angular extent, and ranging from 2.1 to 6.6 pixel per PSF, with an average value of 4.2. The ensuing correlation between the successive radial bins in a given profile is encoded in the correlation matrix. The 120 points sampling applies to the setups D7$_{120}$, P7$_{120}$, and P7. {(All the configurations defined above are summarised  in Table~\ref{t:val})}.

We have compared the individual integrated SZ fluxes one-to-one and the stacked $y$ profiles over the two samples for the various  setups summarised in Table~\ref{t:val}. We chose as reference fluxes the estimation  of  $Y_\textrm{500}$ as derived from the matched multi-filter (MMF) procedure described and used in our first \pact\ paper \citep{agh19}. These fluxes were extracted with the MMF positioned at the \act\ cluster coordinates and with a filter size fixed to  $\theta_\textrm{500}$ for each source. For the profile comparison, we have cross-checked the stacked profile over the whole two samples, adopting as reference the $y$ stacked profiles (and its dispersion envelop) derived for the PLCK62 sample by \citetalias{planck2012-V}.

   \begin{figure*}[!th]
   \centering
   \includegraphics[width=0.5\textwidth]{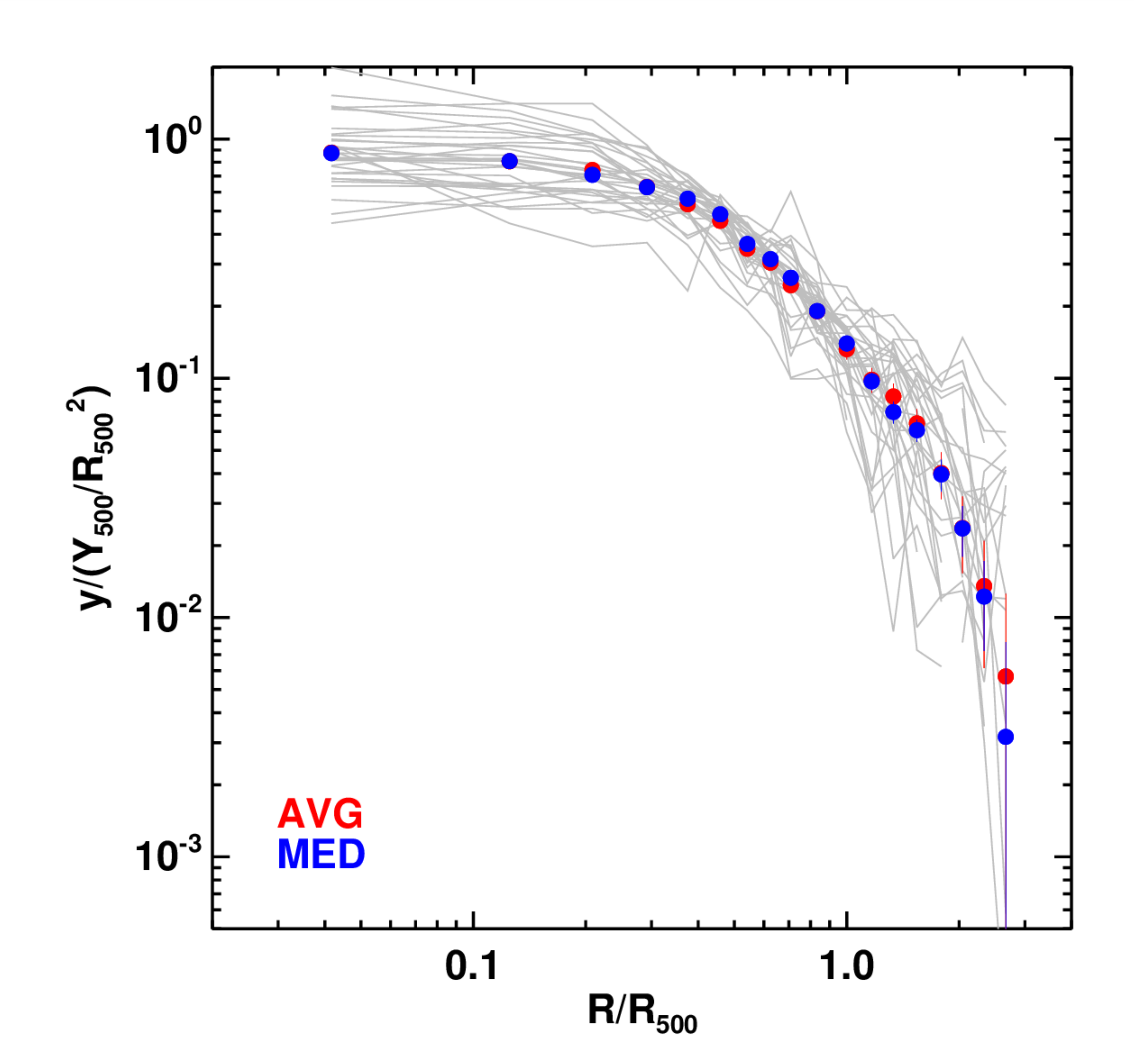}%
   \includegraphics[width=0.5\textwidth]{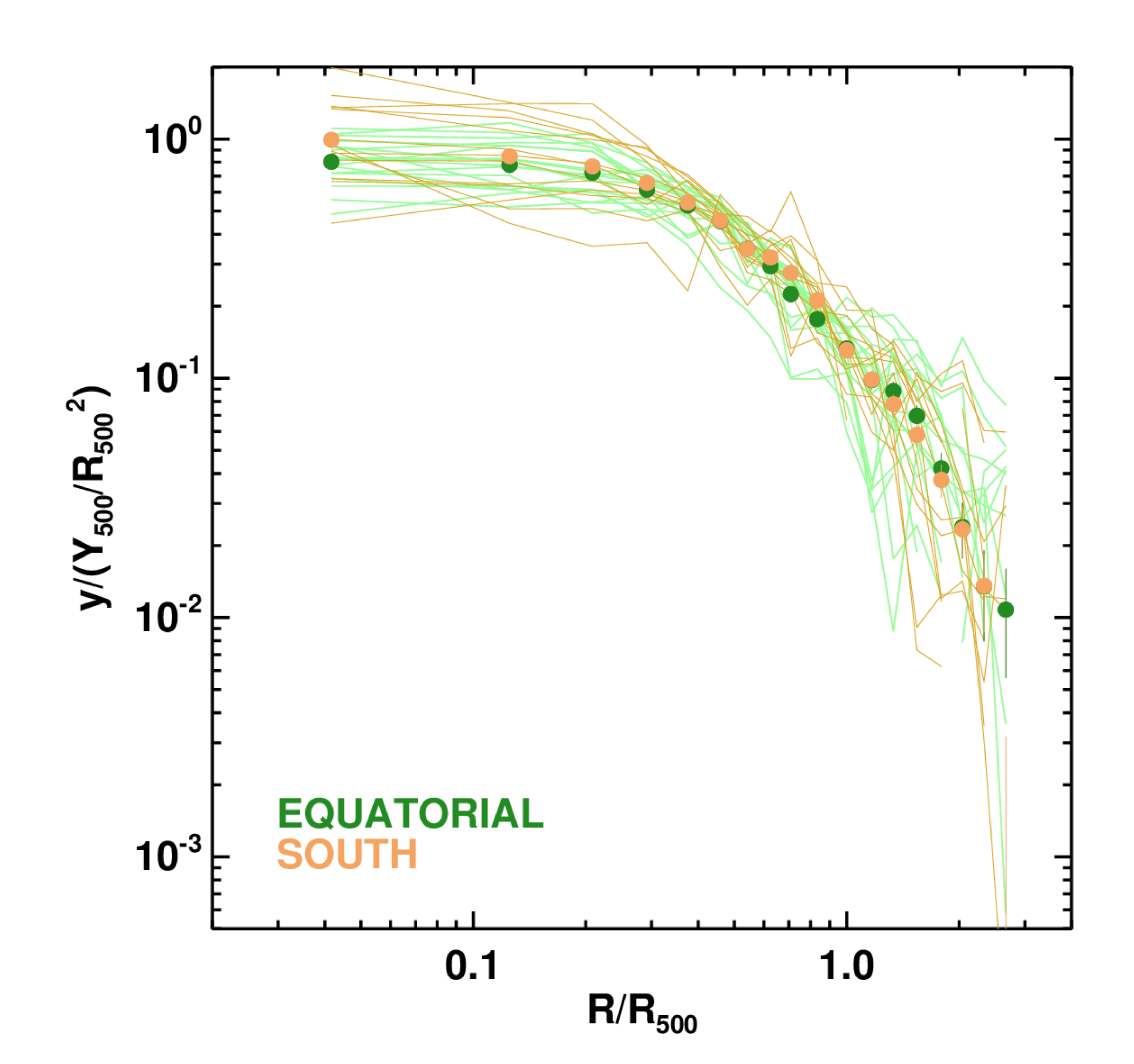}%
      \caption{$y$ profiles for the \asa\ sample.  \emph{(left)} Stacked average (red) and median (blue) profiles. \emph{(right)} Average stacked profiles for the clusters in the equatorial (green) and southern (yellow) strips. The reported errors are correlated and correspond to the square root of the diagonal values of the covariance matrix for the $y$ profile.  Individual profiles are shown as solid lines.}
         \label{f:pr}
   \end{figure*}

%
\subsection{Validation on the \Planck\ $y$-map}
We first compared the integrated SZ fluxes and profiles derived from the first single all-sky \Planck\ survey  \citepalias{planck2012-V} on \psa\ (PIPV) with those derived for the second \Planck\ public release. The latter accounts for more than five co-added all sky surveys (hereafter DR2015). Profiles and fluxes were derived for both all sky $y$-map reconstructed with a  10~arcmin (D10) and 7~arcmin (D7) resolution FWHM, respectively.
The comparisons of fluxes and profiles are shown in Fig.~\ref{f:val}, left-column.
All three flux estimations are consistent with each other and with the \pact\ MMF one. The average ratios to the \pact\ MMF fluxes over the sample are $1.09\pm0.13$,  $1.10\pm0.07$  and  $1.07\pm0.07$ for PIPV, D10 and D7, respectively.  The profiles for the three cases are also fully consistent in shape (the D7 profiles being more peaked is a simple consequence of the smaller PSF which convolves the actual $y$ profile). 
This agreement demonstrates that, for profiles computed with a radial sampling of $\Delta r/R_\textrm{500}=0.25$, there is no bias between the first all-sky survey and the full DR2015 \Planck\ survey, and nor between the flux estimations from the 10~arcmin to 7~arcmin FWHM reconstructed MILCA $y$-map. We thereby adopted the D7 setup for further comparisons.

%
We recomputed over the 7~arcmin FWHM $y$-map, the profiles with a sampling of $\Delta r/R_\textrm{500}=0.08$ and calculated the subsequently associated fluxes (i.e. D7$_{120}$ setup). They are compared in Fig.~\ref{f:val}, middle-column. The profiles are perfectly consistent and so are the fluxes with average ratios to the MMF fluxes of $1.06\pm 0.07$ D7$_{120}$, showing that no bias is introduced by further oversampling the PSF with the corresponding increased bin-to-bin correlation being properly encoded in the correlation matrix.

%
\subsection{Validation on the \pact\ $y$-map}
We switched to  \asa\  and computed profiles and fluxes over the \Planck\ 7~arcmin FWHM DR2015 map with a sampling factor of $\Delta r/R_\textrm{500}=0.08$, that is P7$_{120}$ setup, and over the \pact\ maps with the  same sampling factor, that is the P7 setup. The latest is the nominal setup for the results on the \pact\ map and sample presented in this paper. The comparison of fluxes and profiles (see Fig~\ref{f:val}, right-column) allows us to assess that no bias is introduced due to the difference in sample. The differences between the D7$_{120}$ and P7$_{120}$ stacked profiles reflects the intrinsic difference between the \asa\ and \psa\ samples. As shown in Fig.~\ref{f:sa}, they are populating different regions of the mass and redshift plane. 
The two samples fully overlap in terms of the mass range  (with \psa\ covering a slightly broader range). The  main difference lies in their respective  redshift coverage.
As a consequence the angular sizes of the \asa\ clusters are smaller, hence the increase dilution in the \Planck\ beam explaining the flatter $y$ profile for the P7$_{120}$ setup. As a further consequence, the profile convolution by the beam redistributes power towards larger scales, and explains the slightly  shallower shape at larger radii. 
Unsurprisingly, when switching to the P7 setup, the stacked $y$-profiles for the \asa\ sample as measured from the \pact\ map is more peaked at $r<R_{500}$ than the one obtained from the \Planck\ data only. This is the direct illustration of the differences in smoothing by a PSF of 1.4~arcmin with respect to 10~arcmin (public releases) for \pact\ and \Planck, respectively. However the match in fluxes demonstrates that no bias is introduced {in the integrated SZ flux, $Y_\textrm{500}$,} when switching to the \pact\ maps. The average ratios to the reference MMF values are $1.24\pm 0.27$ and $0.95\pm 0.12$ for P7$_{120}$ and P7 respectively. The results on the \pact\ map and \asa\ sample are further discussed in  the next section.

\section{\pact\  profiles}
\label{s:pr}

Following the validation procedure presented in the previous section, we  consider hereafter the \pact\ results only,  thus we adopted the P7 configuration as defined in Table~\ref{t:val} for the rest of our study.

\subsection{$y$ profiles}
\label{s:pay}

We present in the left panel of Fig.~\ref{f:pr}, the individual  $y$-profiles for the whole \asa\ sample, together with the stacked average and median profiles. The two stacked profiles present no significant differences, hence following  \citepalias{planck2012-V} we adopted the average over the median. The \emph{right} panel shows the colour coded individual and associated stacked profiles for both equatorial and southern strip sub-samples, composed by 18 and 13 clusters, respectively. Within the limits of our sub-sample sizes, we did not find any significant differences between the two, and thereafter consider the \asa\ sample as a whole.

\begin{figure}[!t]
   \centering
   \includegraphics[width=0.5\textwidth]{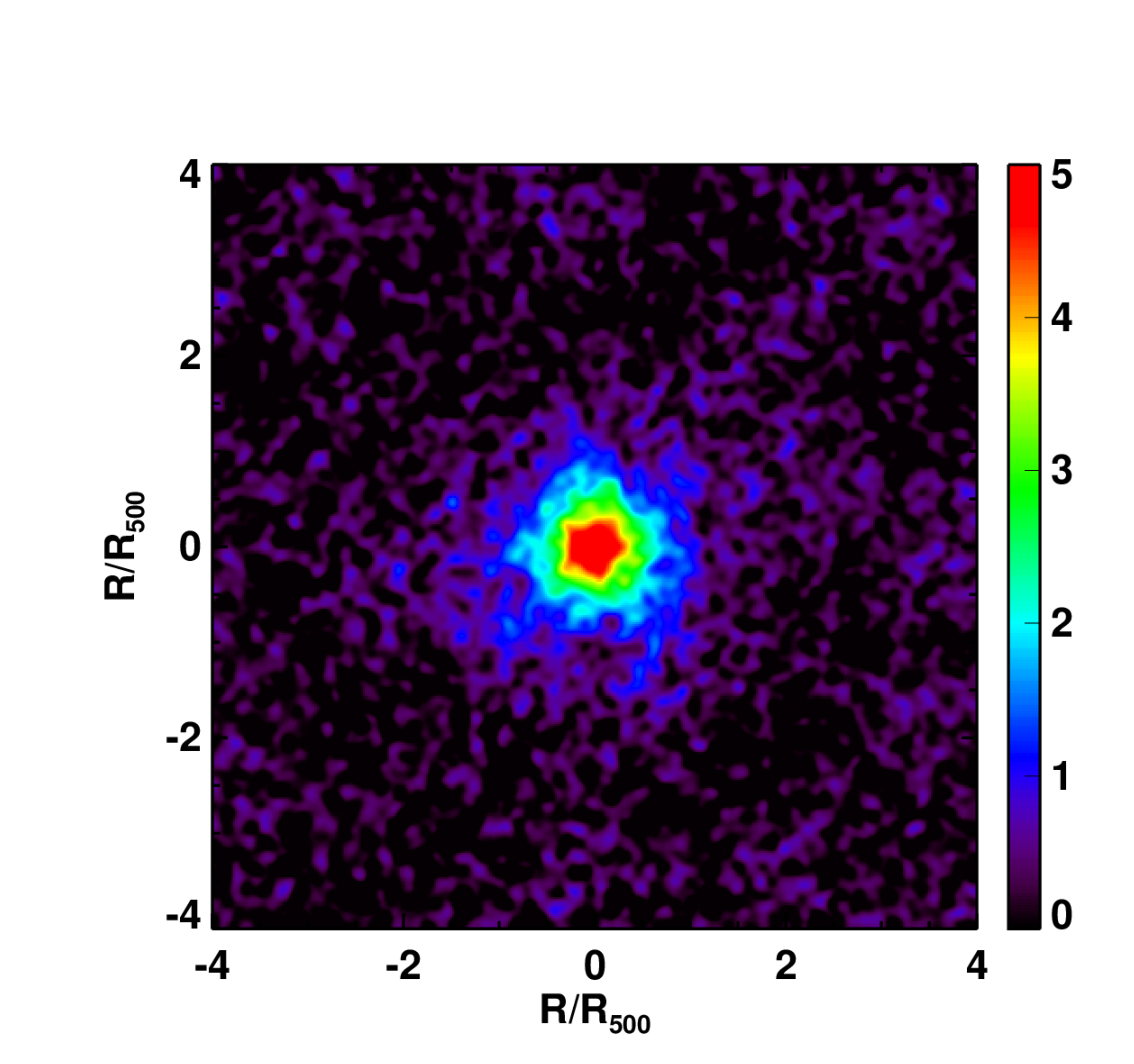}%
      \caption{Stacked scaled $y$ map over the \asa\ sample (size is $8\times  R_\textrm{500}$ on a side, displayed in units of $10^{-5}\times y$). Individual maps are rescaled by $\Phi_i$ (see Sec.~\ref{s:rec}) before averaging, and then multiplied by  $\langle\Phi_i\rangle$.  
      }
         \label{f:im}
   \end{figure}

To give further support to our result, we followed the procedure described by  \citetalias{planck2012-V}, and we stacked the individual $y$ maps across the sample. Each individual map, $m_i$, was rescaled by the factor $\Phi_i$ (see Sec.~\ref{s:rec}) and randomly rotated by 0, 90, 180 or 270$^\circ$ before averaging. The stacked map was finally converted in units of Comptonisation parameter by $\langle\Phi_i\rangle$. A null test map is built from $\sum(-1)^i m_i$. The rms values of this null test map and of the stacked SZ map, outside  $5\times R_\textrm{500}$, are $8.9\times 10^{-6}$ and $4.7 \times 10^{-6}$, respectively. These two rms values are of the same order of magnitude. They are also compatible with the  average value of the stacked error $y$ map, $8.8\times 10^{-6}$. 
The stacked $y$-map is shown in Fig.~\ref{f:im}. It displays a clear SZ signal out to $\sim 2\times R_\textrm{500}$. 
From our stacked average $y$-profile in Fig.~\ref{f:pr}, we detect the SZ signal out to a radius of $\sim 2.5 \times R_\textrm{500}$. This is  slightly less extended than for the \Planck\ sample of  \citetalias{planck2012-V}. This is due to the difference in sample construction, where ours is made of more distant and compact clusters, as illustrated by their respective distribution in the $\theta_\textrm{500}-z$ in Fig.~\ref{f:sa}.

\begin{table}[!tb]
\begin{center}          
     \caption[]{Best fit parameter for the gNFW pressure profile.}
         \label{t:mc}
\begin{tabular}{lccccc}
\hline
\hline
  & $P_0$ & $c_{500}$  & $\alpha$ & $\beta$ & $\gamma$ \\
\noalign{\vskip 2pt\hrule\vskip 2pt}
A10 & $8.40$& $1.18$& $1.05$& $5.49$& $0.31$ \\
P13 & $6.41$& $1.81$& $1.33$& $4.13$& $0.31$ \\
P13$_{NCC}$ & $4.72$& $2.19$& $1.82$& $3.62$ & $0.31$ \\
S13 & $6.41$& $1.81$& $1.33$& $3.67$ & $0.67$ \\
S16 & $9.13$& $1.81$& $1.33$& $6.13$ & $0.31$ \\
\noalign{\vskip 2pt\hrule\vskip 2pt}
P7 & \bf 3.36&    1.18  & \bf 1.08 &  \bf  4.30  &  0.31 \\
& $^{+0.90}_{-0.71}$ & --  & $^{+0.13}_{-0.11}$ & $^{+0.12}_{-0.12}$ & --\\
\noalign{\vskip 2pt\hrule\vskip 2pt}
\end{tabular}
\end{center}
\footnotesize{
\textbf{Note:} A10, P13, S13 and S16 are the parameterisations provided by \citetalias{arn10}, \citetalias{planck2012-V}, \citet{say13,say16}, respectively, and tested against our average stacked profile. P7 corresponds to the best fit parametrisation for this work. Best fit values for free parameters are printed in boldface. { Errors accompanying P7 give the 68\% confidence interval of the marginalised distribution for each parameter.}
}

\end{table}

\subsection{Pressure profile}
\label{s:pap}

From the individual $y$ profiles, we followed  the methodology presented in \citetalias{planck2012-V} to reconstruct the 3D pressure profiles of each of our clusters. We then stacked them into a normalised average profile and also computed the associated stacked covariance matrix (details of the formalism are provided in Sec.~\ref{s:rec}, see Eq.~\ref{e:norm}). 

\begin{figure*}[!t]
   \centering
   \includegraphics[width=0.5\textwidth]{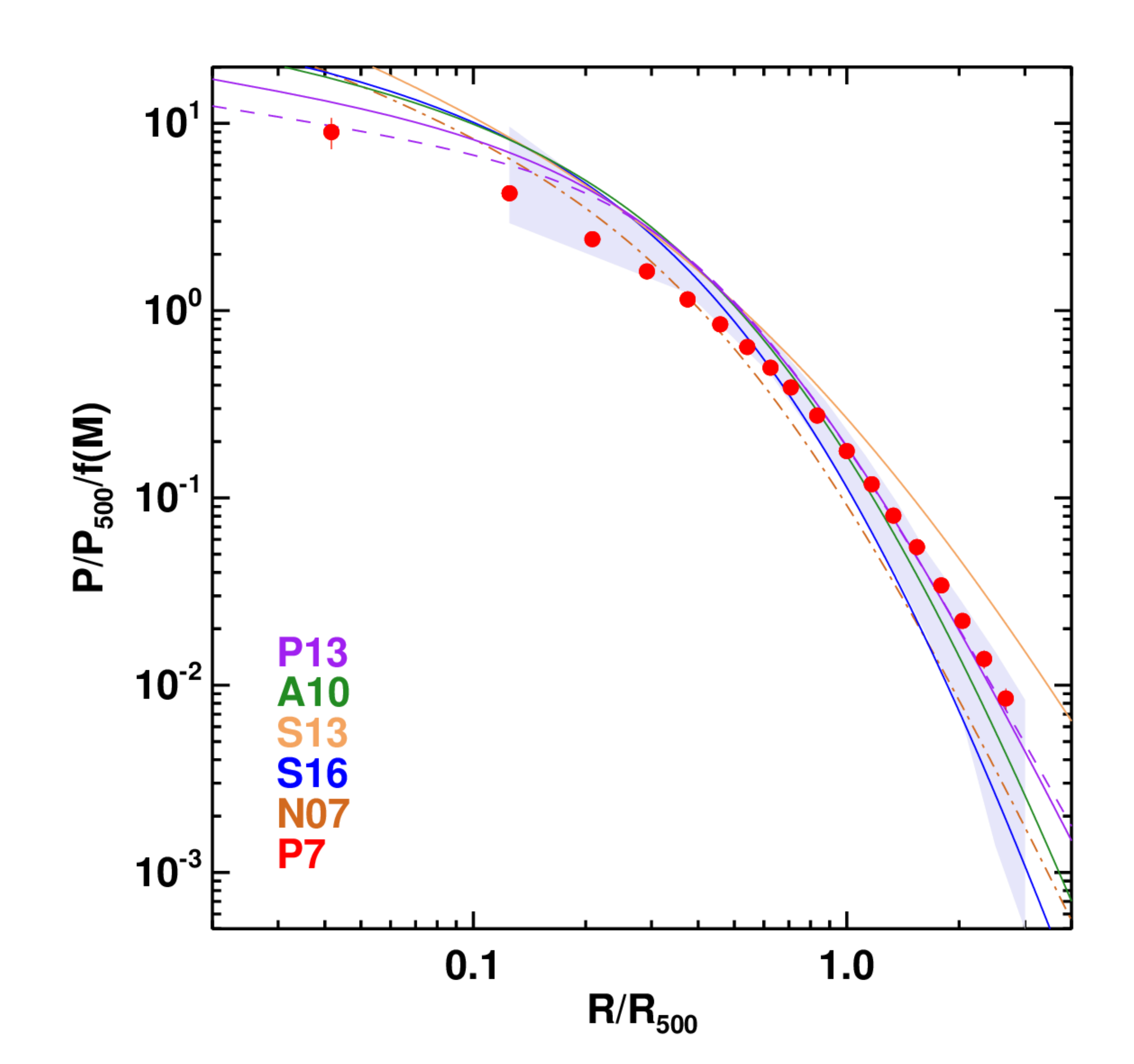}%
    \includegraphics[width=0.5\textwidth]{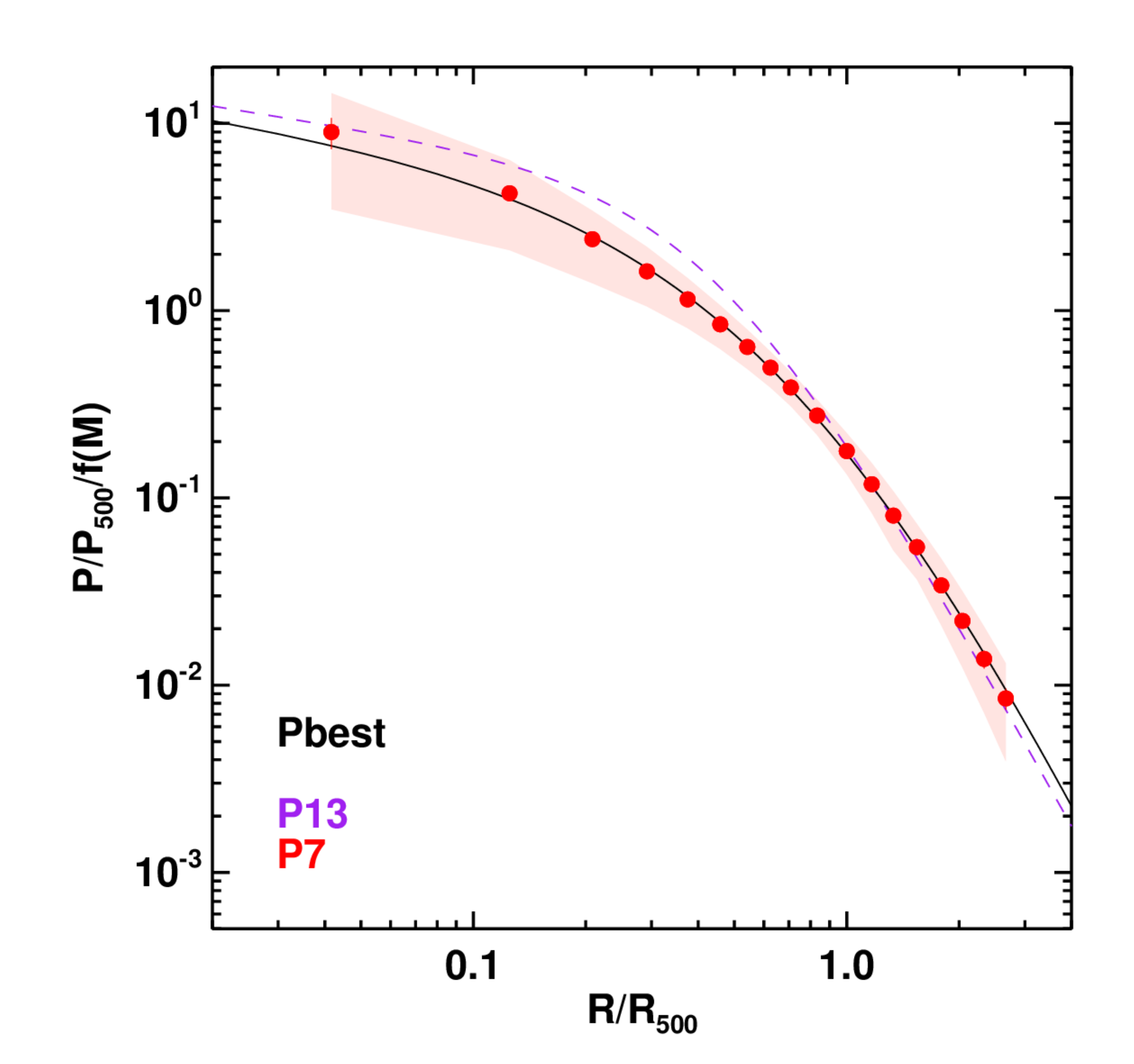}%
     \caption{Stacked average pressure profiles over the \asa\ sample (red points). The reported errors bars on the data points are correlated and correspond to the square root of the diagonal valued of the covariance matrix for the pressure profile \emph{(left)} Comparison to published profiles over different samples by (\citetalias{arn10}; \citetalias{planck2012-V}; \citealt{say13,say16}) are overlaid as green, purple, yellow and blue lines and labelled as A10,  P13, S13 and S16, respectively. {The brown doted-dashed line, labelled N07, shows the original parametrisation from \citet{nag07}.}For the  \citetalias{planck2012-V} profiles, the solid  and dashed lines correspond to the best fit to the whole \psa\ sample and the non cool-core sub-sample, respectively. The purple shaded area picture the dispersion of the stacked \Planck\ profiles for the \psa\ sample \citepalias[as published by][their figure~4]{planck2012-V}. \emph{(right)} Best fit of our data to a gNFW pressure distribution (solid black line -- see Table~\ref{t:mc} and Sec.~\ref{s:pap}). The red shaded area shows the dispersion of the stacked profiles for the \asa\ sample. The dashed purple solid line is identical to the left panel.}
         \label{f:ppr}
   \end{figure*}

The derived stacked pressure profile for our sample is shown in Fig.~\ref{f:ppr} (right panel) and compared to previous results from different samples (left panel).
Altogether, it is in perfect agreement with the profile derived from the  \psa\ sample by  \citetalias{planck2012-V} within the dispersion of both samples. We recall that the \asa\ and \psa\ samples contain 31 and 62 clusters, respectively. The outer slopes for both samples are in good agreement, whereas the inner part  (i.e. $r<R_\textrm{500}$) differs slightly,  our \pact\ profile more shallower  than the \psa\ non-cool core sub-sample.

\begin{figure*}[!t]
   \includegraphics[width=0.33\textwidth]{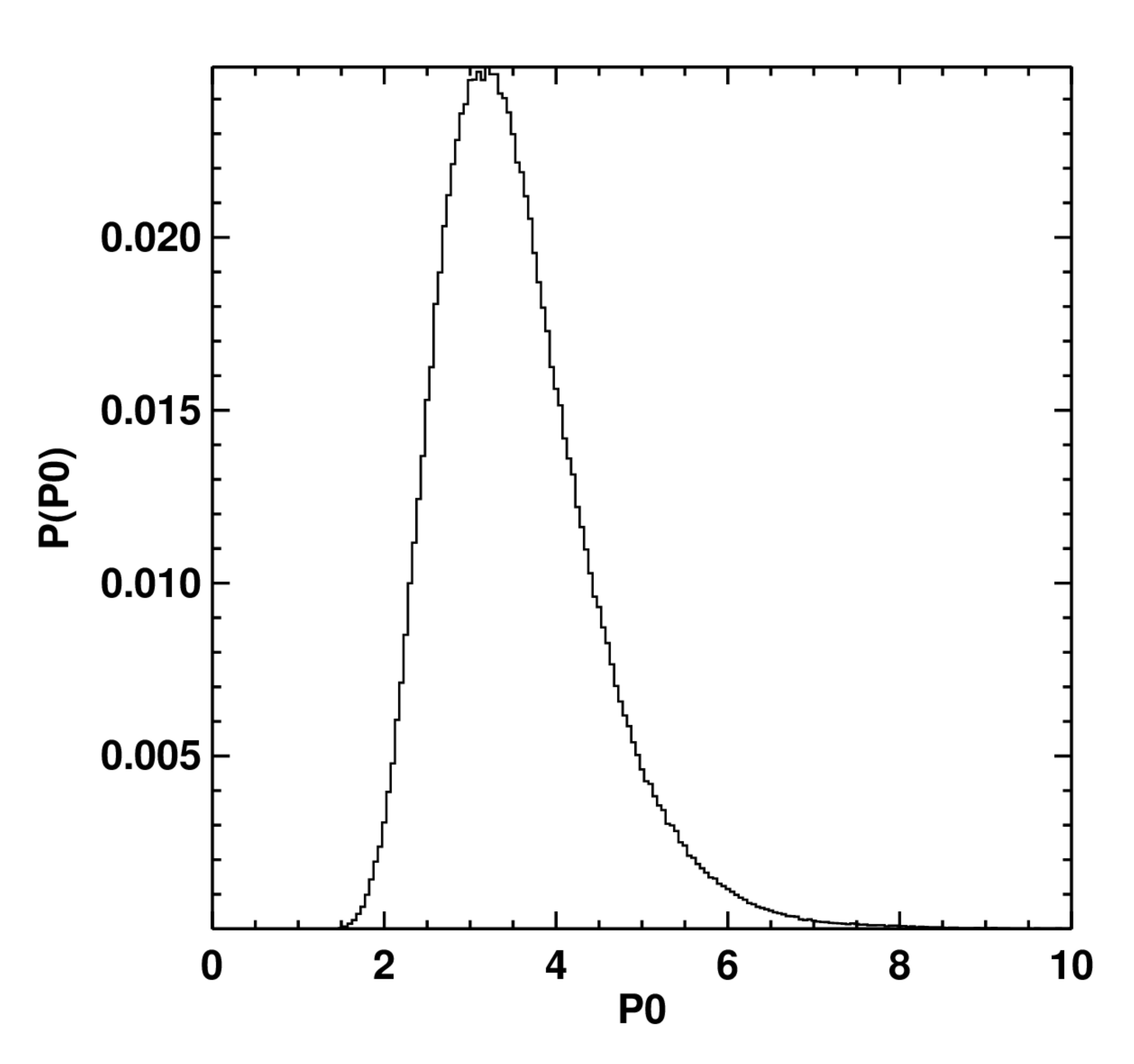}\\
   \includegraphics[width=0.33\textwidth]{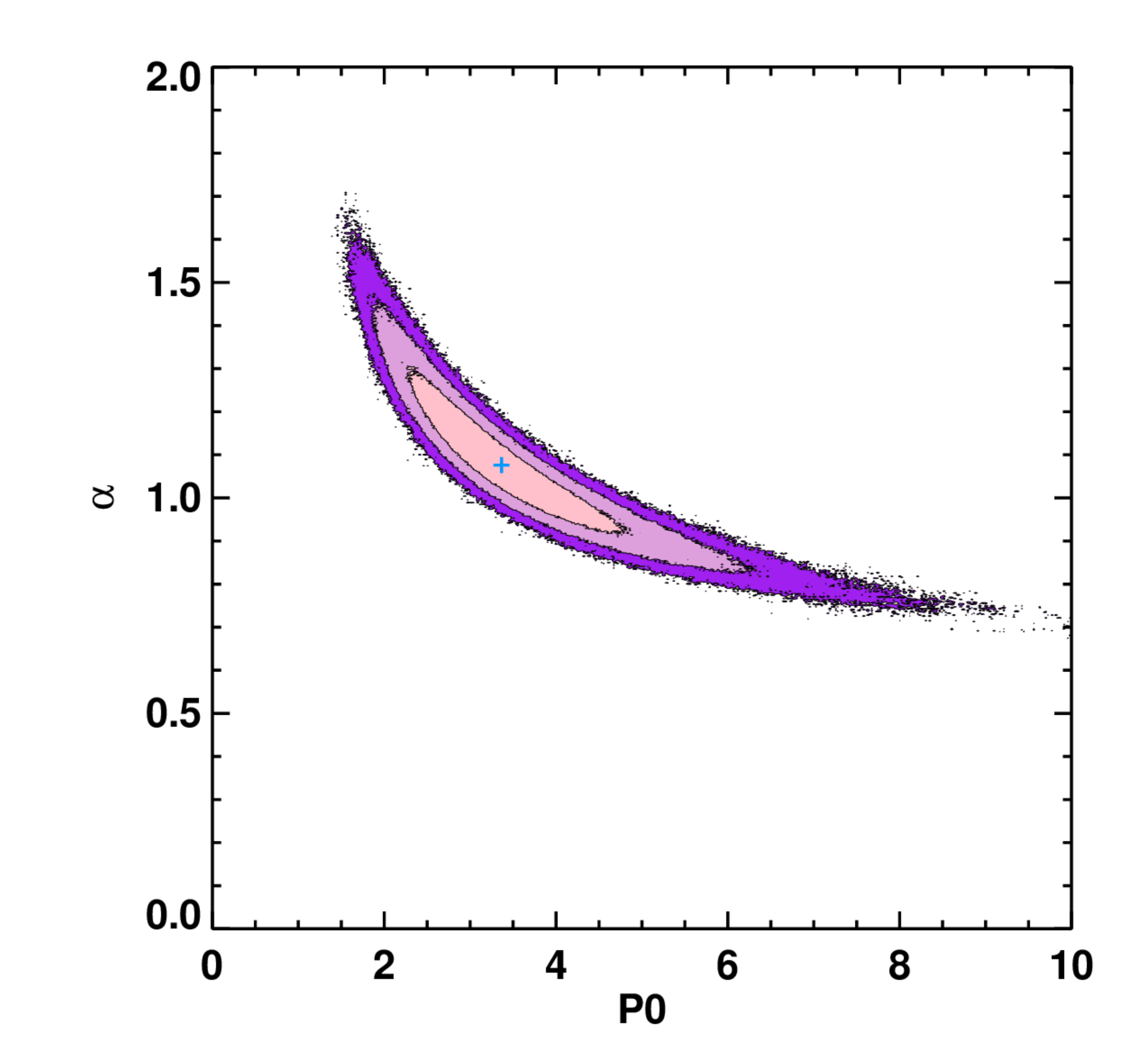}%
   \includegraphics[width=0.33\textwidth]{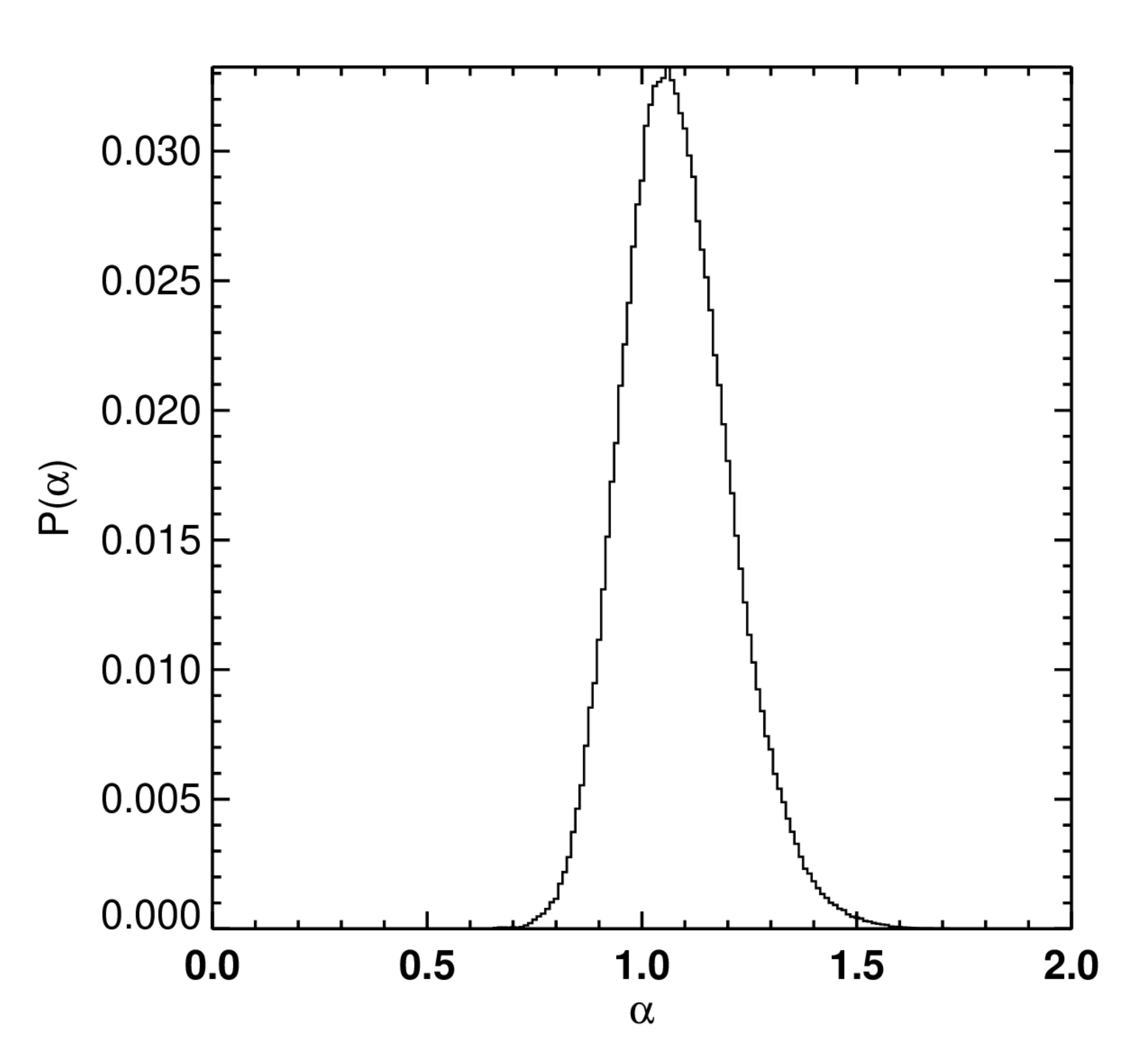}\\
   \includegraphics[width=0.33\textwidth]{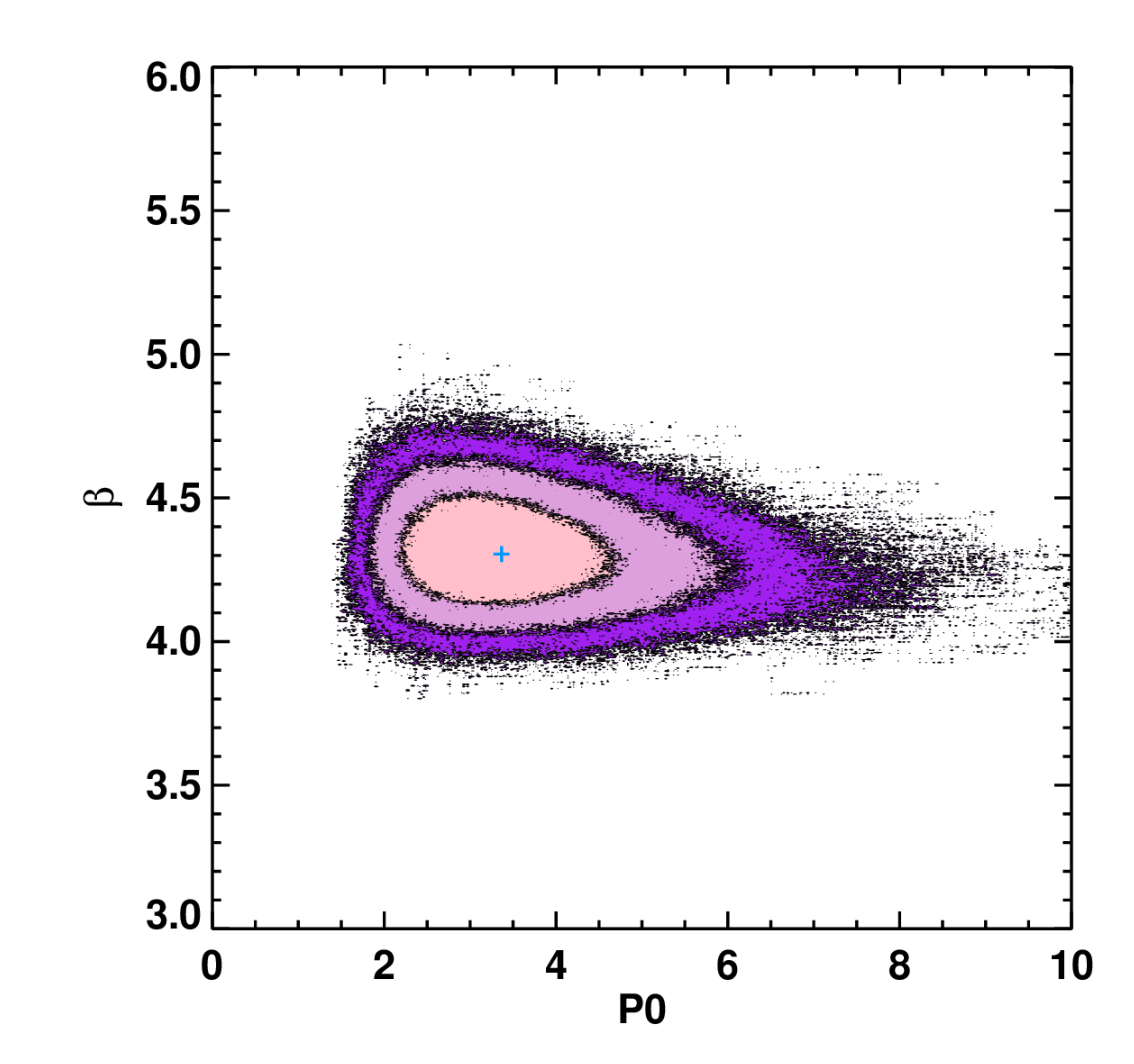}%
   \includegraphics[width=0.33\textwidth]{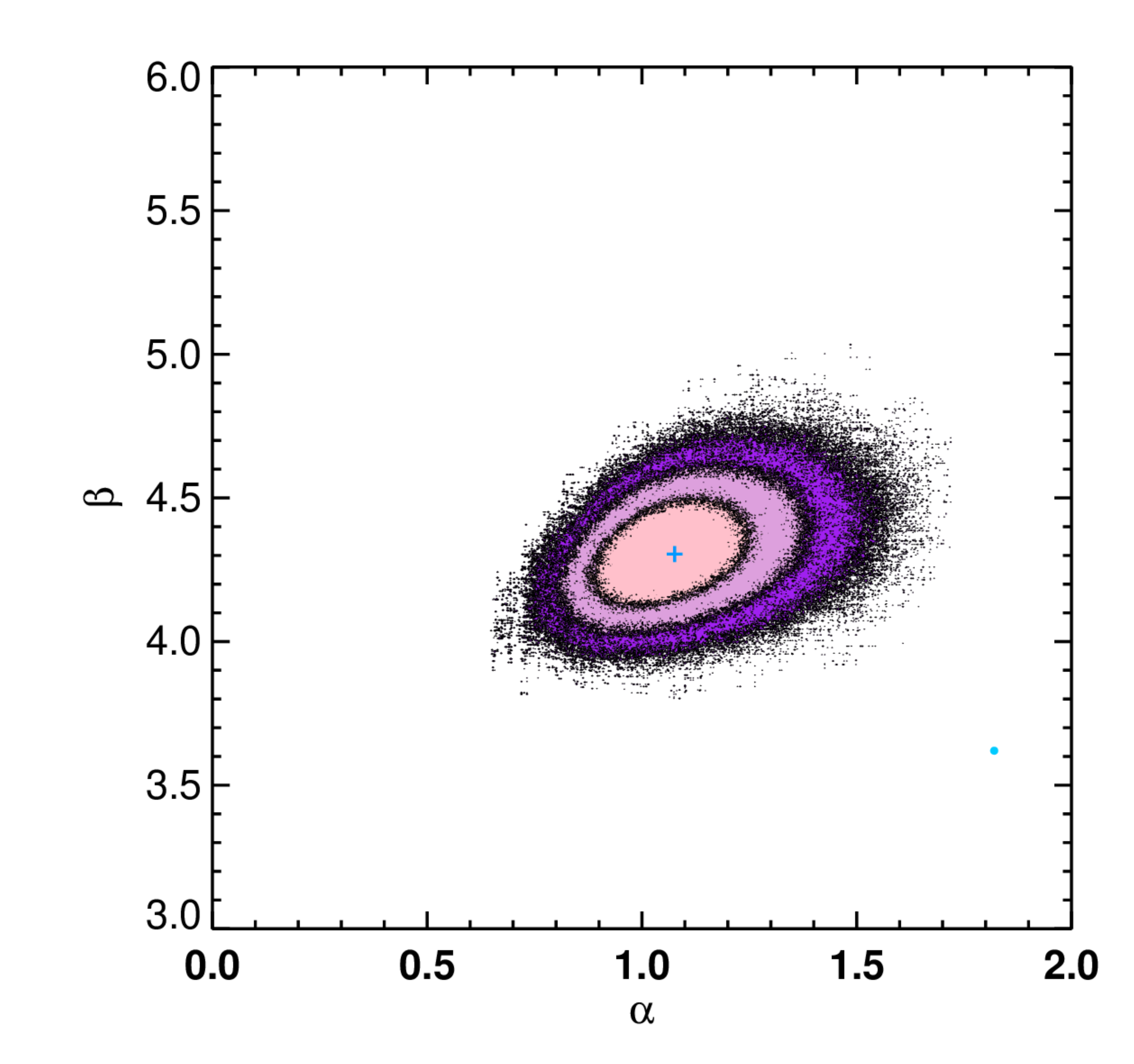}%
   \includegraphics[width=0.33\textwidth]{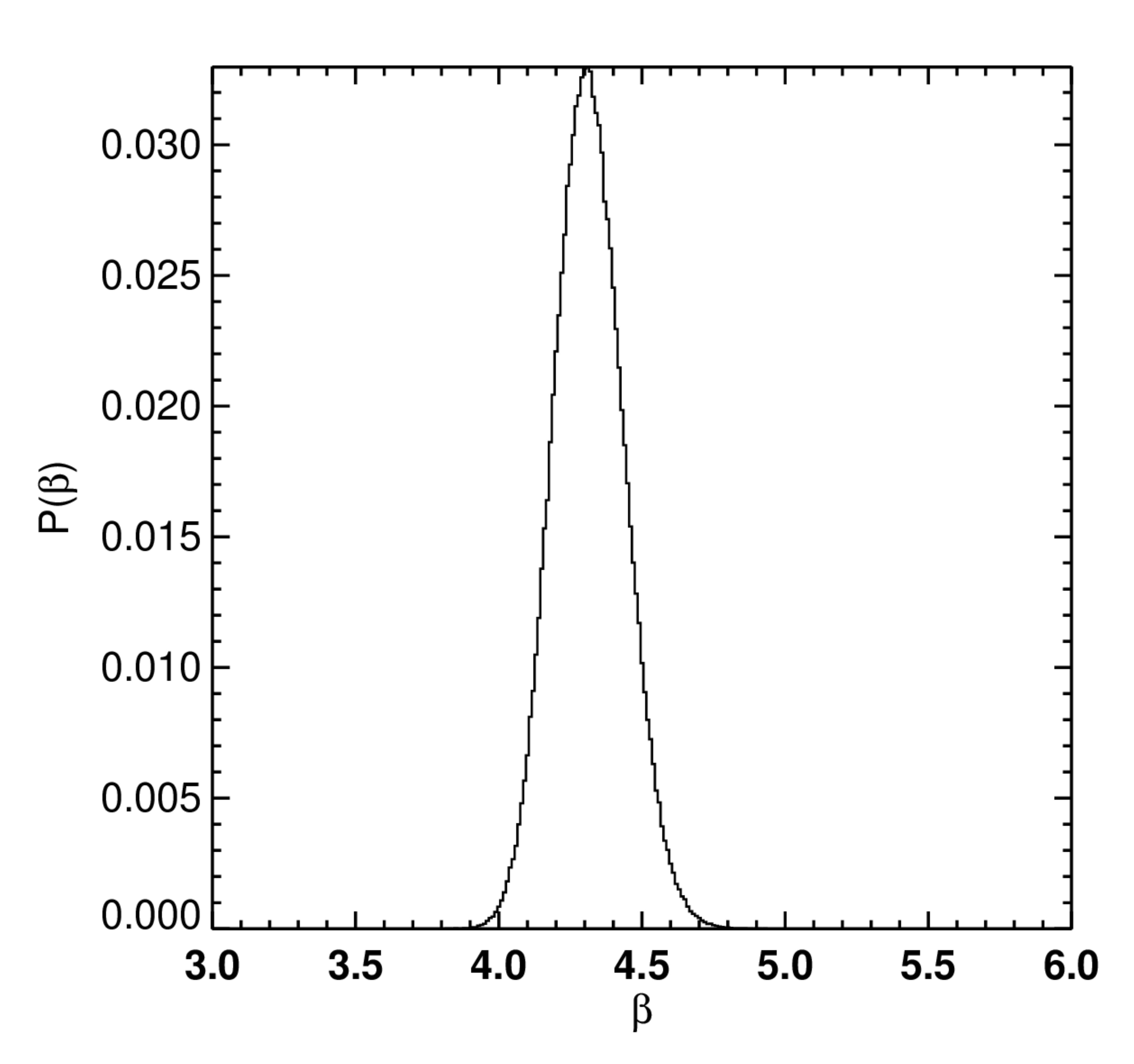}%
    \caption{{Posterior probability distributions and heat maps corresponding to our  MCMC fit to a gGNFW pressure profile (see Sec.~\ref{s:pap}). The blue crosses in the 2D heat maps show the optimal solution  reported in Table~\ref{t:mc}. The colour filled area show the locus of the 68, 95 and 99.7\% confidence levels, respectively.}
    }
         \label{f:cor}
   \end{figure*}

We fitted our stacked mean pressure profile to a gNFW model \citep{nag07} with a Monte Carlo Markov chain{, making use of an implementation based on the Metropolis–Hastings algorithm \citep{hur96,bra05,zob11}}. We accounted in the fit for the correlation between our points through {the covariance matrix of the pressure profile,} and of the dispersion on the mean across our sample. {The latter is quadratically added to the diagonal elements of the profile covariance matrix}. Following \citetalias{planck2012-V} and from the previous consideration we performed two fits with four and three free parameters, respectively. In both cases, {since we lack the spatial resolution to resolve the inner profile, we fixed the slope to $\gamma=0.31$, the best fit value from \citetalias{arn10}. We note that the joint MUSTANG-1 and Bolocam study by \citet{rom17} also converge towards this value. } The four parameter fit let $P_0$, $c_\textrm{500}$, $\alpha$, and $\beta$ be free with uniform prior intervals of [0.5,20], [1,3], [0.1,4], and [0,8], respectively. The three parameter fit uses the same configuration, but the concentration parameter  is fixed to the best fit value of  \citetalias{arn10}, that is $c_\textrm{500}=1.18$. 
{Each MCMC fit was run with a 100 chains ending up with a number of iterations of $\sim 30000$ each in the converged final chain. The fit processed is assessed on the basis of the likelihood logarithm, that is $-\chi^2/2$ and  $\chi^2 = (P-M)^T C^{-1} (P-M)$, where $P$, $M$ and $C$ are the observed profile, the model profile and the profile covariance matrix, respectively.} 
The fit with four free parameters leads to a solution where $c_\textrm{500}$ hits  the lower boundary of the its prior  interval. Letting this parameter run towards smaller value leads to a catastrophic degeneracy with the outer slope $\beta$. The scale radius $r_s=R_\textrm{500}/c_\textrm{500}$ becomes very large, inducing a similarly large value for the external slope $\beta$, hence to a quite unphysical solution. We have thus preferred  the fit with three free parameters, where the scale parameter is fixed. The results of the three-parameter fit are gathered in Table~\ref{t:mc} together with the same from previous works, and displayed in Fig.~\ref{f:ppr}. {The heat map for each pair of free parameters and the associated posterior probabilities for $P_0$, $\alpha$, and $\beta$ are shown on Fig.~\ref{f:cor}. The 68\% confidence-level errors associated with the parameters are derived from the posterior probabilities and reported in Table~\ref{t:mc}.}
{For our best fit with three free parameters, we find a $\chi^2=2.19$. 
We attribute the relatively low value of the $\chi^2$ to the fact of considering for the uncertainties of the stacked profile both the statistical uncertainties (propagated from the individual profiles) and the dispersion across the sample as discussed above. This  might  overestimate the actual uncertainties and as so artificially reduce the $\chi^2$.
Conversely, the purely statistical error conveyed by the covariance matrix lead to a $\chi^2$ of $43.5$ for our 18 data points profile. In this case the error is likely underestimated possibly due to some non-Gaussian correlated noise component in the $y$-map. In order to further assess the quality of our derived best parametrisation, we computed the associated F-test to our best fit against our pressure profile data points, and derived  a significance of $0.49$ denoting the goodness of our fit.}

\section{Discussion and conclusion}
\label{s:dis}

 From our Fig.~\ref{f:ppr}, it is clear that the derived average stacked pressure profile for our \asa\ sample is  consistent at large scales in external slope  and normalisation with previous published works from distinct samples (i.e. \citetalias{arn10}, \citetalias{planck2012-V}, \citealt{say13, say16}). Within  $\sim 0.5\times R_\textrm{500}$ our profile is slightly flatter, although the differences remain within  the dispersions of these respective samples.

{As emphasised by \citet{mro19}, the point of a universal pressure  profile is in its shape more than in the intrinsic values of its parameters. In our case, we note the very consistent shape of the outer parts of our profile with that derived from \citetalias{planck2012-V}.  We are more marginally compatible with the \citetalias{say13} and \citetalias{say16}, which are bracketing our dispersion. Within this dispersion, we are also compatible at large radii with the \citetalias{arn10} profile, hence with the theoretical predictions from numerical simulations by \citet{bor04,nag07,pif08}, against  which it is constrained beyond $R_\textrm{500}$. The outer profile is also consistent (similar to \citetalias{planck2012-V}) with the predictions from \citep{bat12,gup17}. This result denotes that for both \citetalias{planck2012-V} and our profile, the large scales are constrained by the \Planck\ measurements, with no obvious signature due to the differences in sample composition. The radial range  [0.1,1]~$R_\textrm{500}$ presents the most differences with previously published profiles.  As the \citetalias{planck2012-V} profile is derived from a joint \Planck\ and \xmm\ analysis, the inner parts are strongly constrained by the X-ray data (see their Fig~4, left panel). A fit  performed on the \Planck\ profile only, may have lead to a shallower profile. At the same time we are mainly consistent within this radial range and the respective dispersions with the \citetalias{say16} profile.  Conversely, as shown on the left panel of Fig.~\ref{f:ppr}, the comparison to the original parametrisation $[P_0 h_{70}^{2/3},c_\textrm{500},\alpha,\beta,\gamma]=[3.3, 1.8,1.3,4.3,0.7]$ published by \citep{nag07} normalised against \emph{Chandra} observations of relaxed clusters  \citep{vik06} is mostly consistent with our profile within this inner radial range. The same stands for an update of this parametrisation, $[\alpha,\beta,\gamma]=[0.9,5.0,0.4]$ presented in \citet{mro09}.}
{The shallower shape of our profile in the central parts is likely due to the fainter and more compact nature of our clusters relative to those of \citetalias{planck2012-V} and  \citetalias{say16}.}

{\citet{say16} noted that the differences found in pressure profile analyses originate from various possible factors: the sample definition and selection, the biases intrinsic to instruments, and methods for data analysis. The comparison to the \psa\ sample minimises  the possible sources of these differences as we adopted the same data analysis methodology. While the instrumental setups differ,  the approach for the reconstruction of the \pact\ y-map is similar to that of the \Planck\ survey. We refer here to \citet{agh19} for the extensive validation of this dataset with respect to both \act\ and \Planck. For these two samples we  attribute the differences in the mean pressure profile to the difference in composition of the \asa\ and \psa\ samples}.  As shown in Fig.~\ref{f:sa}, the \asa\ and \psa\ selections do sample two different regions of the $M-z$ plane (and subsequently the $M-\theta$ and $\theta-z$ planes). Their direct comparison is  not straightforward. Their respective selection function is different{, and in both cases, neither quantified nor easy to apprehend. The \psa\ sample is composed of  62 clusters with $S/N>6$ from the first \Planck\ all-sky survey. Our \asa\ sample, though of reasonable statistically size, is two times smaller with 31 clusters based on the union of the full \Planck\ survey (i.e. more than 5 all-sky passes) and the \act-MBAC survey with S/N  going down to 4 in the \Planck\ PSZ2 catalogue (see Table~\ref{t:sa}).}
As a simple verification, if we only consider clusters in the common box interval of $0.15<z<0.45$ and $5\times 10^{14}<M_\textrm{500}<1\times 10^{15}$~M$_\odot$ between the \psa\ and \asa\ samples, we obtained averaged values for the SZ flux (i.e. PSZ2 integrated Comptonisation parameter $Y_\textrm{500}$), of $2.4\times 10^{-3}$ and $1.3\times 10^{-3}$~arcmin$^2$, from 25 and 20 clusters, respectively. The latter is on average fainter by a factor of $\sim2$. {This discrepancy is increased to a factor of $\sim8$ when $Y_\textrm{500}$ are compared in units of Mpc$^2$ . This renders the  detection of the SZ signal more difficult in the case of our \asa\ sample. As they are also more compact on average, the reconstruction of their 3D pressure profile is more prone to effects such as smoothing at large scales by the \Planck\ beam, and thus potential biases from the regularised deconvolution and deprojection process.}

{Our \asa\ sample, as with all the samples used in previous published works on pressure profiles from SZ observations, has been assembled on a best-effort basis given specific observational constraints and working contexts. For instance, the \psa\ sample is SZ selected, but is constrained by the \xmm available archive data at the time of \citetalias{planck2012-V}'s publication. Our \asa\ sample is fully SZ selected but constrained on the basis of the \pact\ construction and coverage.  The Bolocam sample used in \citetalias{say13} and \citetalias{say16} was assembled on the basis of X-ray coverage by \emph{Chandra} from the CLASH \citep{pos12} and MACS samples \citep{ebl07}. 
Thought, not actually SZ selected the SPT sample, followed-up through a very large \chandra\ programme, has led to several key results \citep[e.g.][]{mcd13, mcd16}.
None of these  is representative of the cluster population in its sampling of the mass and redshift space. The only representative sample to which we compare to is the \rexcess\ sample \citep{boe07} from which the  \citetalias{arn10} universal pressure profile is parametrised. However this is an X-ray selected sample.  Quantifying their differences and trying to promote one as a reference versus the others is therefore a complex and risky task. This limitation makes any extensive discussion on the physical meaning of the  differences in the central parts of our mean pressure profile to others quite speculative at this stage. For instance, if we consider the evolution with redshift of the intrinsic SZ flux, $Y_\textrm{500}$, at a given mass, the self-similar evolution  is expected to be proportional to $E(z)^{2/3}=[(1+z)^3\times \Omega_m +\Omega_\Lambda]^{1/3}$. At the average redshift value of the \asa\ and \psa\  samples, that is $0.33\pm0.11$ versus $0.17\pm0.11$, respectively, we expect  an average evolution of $\sim 6$\%. Accounting for the dispersion in redshift over the two samples, this expectation is uncertain by $\sim 12$\%. The likely differences in population sampling for the two samples, the associated dispersion of each sample in the $M-z$ plane, and their proximity in redshift convolved by their underlying selection function prevent us from drawing any serious constraints on the evolution of the pressure profile.}

The advantage provided by the combination of the \Planck\ and \act\ data lies on the combination of the \act\  higher spatial resolution and the \Planck\ large-scale coverage. It provides a more accurate description of the SZ signal from the central parts to the outskirts of individual  clusters. 
Towards the cluster centre our constraints stretch down to $r=0.04\times R_\textrm{500}$ (see Fig.~\ref{f:ppr}). 
This can be compared to the SZ-only reach of the \citetalias{planck2012-V} pressure profiles for the \psa\ sample, which is $0.125\times R_\textrm{500}$ (see the red envelop in Fig. ~\ref{f:ppr}). With our adopted parametrisation (see Table~\ref{t:val}) this is only illustrated by an extra central point in the reconstructed  3D pressure profile with respect to \citetalias{planck2012-V}. This is however an improvement  by more than a factor 3 in the central radial reach (obtained with a sample two times smaller). Our radial coverage towards the centre falls short  of the \citetalias{planck2012-V} joint \xmm\ X-ray and \Planck\  SZ  profile, which goes down in radius by an extra factor of 2,  to $0.02\times R_\textrm{500}$. Such a resolution is  achievable in SZ alone, making use of  high resolution facilities such as \mustang-2 and \nika-2 \citep[e.g.][]{rup18,rom20}, though assembling data for a sample as large or larger than our \asa\ sample with these two facilities would require a very large amount of time (e.g. the \nika-2 SZ guaranteed-time programme of 300~hours for 45 clusters, \citealt{may20}).  
~\\

{In conclusion, we have demonstrated the self consistency of two sets of SZ data by combining \Planck\ and \act\ data  to reconstruct the ICM gas pressure distribution. We note that, as for previous works, the non-representative nature of our sample limits its broader applicability, beyond the basic comparisons given here. An unbiased sample at low-to-intermediate redshifts is needed  to serve as a reference for the community. Such a sample will have to be a carefully selected sample, based on large SZ catalogues of clusters \citep[e.g.][]{planck2013-p05a,planck2014-a36,ble20,hil20}.The CHEX-MATE sample} from the \xmm heritage class programme '\emph{Witnessing the culmination of structure formation in the Universe}' fulfils this requirement by construction. \citep{arn20}. SZ selected from the \Planck\ survey, it gathers 118 clusters with $S/N_\textrm{PSZ2}>6$ and shall be fully covered by deep \xmm observations. The combination of the \xmm and \Planck\ data following the work from \citetalias{planck2012-V} will  provide a precise joint X-ray and SZ view of the ICM properties over this sample. This will provide a solid base  to investigate  changes in the pressure distribution with mass and redshift. Indeed, such variations impact the detection of clusters and the modelling of the SZ spectrum, as both rely on knowing the spatial distribution for the thermal pressure.
Furthermore, as demonstrated in the present work, the combination with higher resolution SZ data constitutes a key improvement for its physical characterisation. Pointed observations with facilities such as \mustang-2 and \nika-2 will be an asset, though the largest coverage of the aforementioned sample shall be achieved by the \emph{Adv}\act\ survey,  { for which frequency maps over $\sim 18000$~sq. deg. of the sky were recently publicly released \citep{nae20}}.

\begin{acknowledgements}
This research was performed in the context of the ISSI international team project 'SZ clusters in the \Planck\ era' (P.I. N. Aghanim \& M. Douspis). We thank the anonymous referee. We thank Matthew Hasselfield for his participation to the early phase of this work. The authors acknowledge partial funding from the DIM-ACAV, the Agence Nationale de la Recherche under grant ANR-11-BS56-015, and PNCG/INSU/CNRS. The research leading to these results has received funding from the European Research Council under the H2020 Programme ERC grant agreement no 695561. JMD acknowledges the support of project PGC2018-101814-B-100 (MCIU/AEI/MINECO/FEDER, UE) Ministerio de Ciencia, Investigaci\'on y Universidades.  This project was funded by the Agencia Estatal de Investigaci\'on, Unidad de Excelencia Mar\'ia de Maeztu, ref. MDM-2017-0765 DC acknowledges support from the South African Radio Astronomy Observatory, which is a facility of the National Research Foundation, an agency of the Department of Science and Technology. The development of \Planck\ has been supported by: ESA; CNES and CNRS/INSU-IN2P3-INP (France); ASI, CNR, and INAF (Italy); NASA and DoE (USA); STFC and UKSA (UK); CSIC, MICINN, JA and RES (Spain); Tekes, AoF and CSC (Finland); DLR and MPG (Germany); CSA (Canada); DTU Space (Denmark); SER/SSO (Switzerland); RCN (Norway); SFI (Ireland); FCT/MCTES (Portugal); and PRACE (EU). This research has made use of the following databases: the NED and IRSA databases, operated by the Jet Propulsion Laboratory, California Institute of Technology, under contract with the NASA; SIMBAD, operated at CDS, Strasbourg, France; SZ cluster database (szcluster-db.ias.u- psud.fr) operated by Integrated Data and Operation Centre (IDOC) operated by IAS under contract with CNES and CNRS. This research made use of Astropy, the community-developed core Python package.
\end{acknowledgements}

%
%
\bibliographystyle{aa.bst}
\bibliography{pppact,Planck_bib}

\end{document}